\documentclass[a4paper,10pt,openany]{article}
\usepackage{pst-3d}
\usepackage{pst-3dplot}
\usepackage{etex}
\usepackage[utf8]{inputenc}
\usepackage[english]{babel}
\usepackage{amsmath,amssymb,amsthm,mathrsfs,amsfonts,dsfont}
\usepackage{graphicx}
\usepackage[footnotesize, skip=0pt]{caption}
\usepackage{epigraph}
\usepackage{indentfirst}
\usepackage{booktabs}
\usepackage{fancyhdr}
\usepackage{vmargin}
\usepackage{feynmf}
\usepackage{yfonts}
\usepackage{lscape}
\usepackage{subfig}
\usepackage{textcomp}
\usepackage{pstricks}
\usepackage[metapost]{mfpic}
\usepackage{fancybox}
\usepackage{slashed}
\usepackage[verbose]{wrapfig}
\usepackage{pifont}
\usepackage{keystroke}
\usepackage{appendix}
\usepackage{enumitem}
\usepackage{dyntree}
\usepackage{array}
\usepackage{colortbl}
\usepackage{alltt}
\usepackage{boxedminipage}
\usepackage{tikz}
\usepackage{calc}
\usepackage{subfloat}
\usepackage{multirow}
\usepackage{cmll}
\usepackage{multicol}
\definecolor{color1}{RGB}{204,0,51}
\definecolor{color2}{RGB}{159,182,205}
\usepackage{bookmark}

\usepackage{verbatim}
\usepackage[vcentermath]{youngtab}
\usepackage{young}
\usepackage{pst-grad} 
\usepackage{pst-plot} 
\usepackage{transparent}
\usepackage{color,soul}
\usepackage{cancel}
\usepackage{bm}
\usepackage{pifont}
\usepackage{placeins}
\usepackage{mathtools}
\usepackage{rotating}
\usepackage[refpage]{nomencl}
\usepackage{hhline}
\usepackage{marginnote}
\usepackage{upgreek}
\usepackage{stackengine}
\usepackage{setspace}
\usepackage{ytableau}
\usepackage{listings}  
\usepackage{calligra}
\usepackage{hyperref}
\usepackage{afterpage}
\usepackage{cite}
\usepackage{cleveref}
\usepackage{makecell}
\usepackage{accents}
\usepackage{breqn}
\usepackage{environ}

\newcommand{\hepth}[1]{{\tt
\href{http://www.arXiv.org/abs/hep-th/#1}{hep-th/#1}}}

\newcommand{\arxiv}[1]{{ arXiv: \href{http://www.arXiv.org/abs/#1}{#1}}}
\newcommand{\doi}[1]{{{\scriptsize DOI:} \href{http://dx.doi.org/#1}{#1}}}

\allowdisplaybreaks
\lstdefinestyle{customc}{
  belowcaptionskip=1\baselineskip,
  breaklines=true,
  frame=L,
  xleftmargin=\parindent,
  language=C,
  showstringspaces=false,
  basicstyle=\footnotesize\ttfamily,
  keywordstyle=\bfseries\color{green!40!black},
  commentstyle=\itshape\color{purple!40!black},
  identifierstyle=\color{blue},
  stringstyle=\color{red217!80!black},
}

\lstdefinestyle{customasm}{
  belowcaptionskip=1\baselineskip,
  frame=L,
  xleftmargin=\parindent,
  language=[x86masm]Assembler,
  basicstyle=\footnotesize\ttfamily,
  commentstyle=\itshape\color{purple!40!black},
}

\sethlcolor{yellow}
 \definecolor{red217}{RGB}{217,0,0}
\definecolor{boh}{RGB}{79,47,79}

\hypersetup{
linkbordercolor={red}
}

\hypersetup{
    bookmarks=false,         
    unicode=false,          
    pdftoolbar=true,        
    pdfmenubar=true,        
    pdffitwindow=false,     
    pdfstartview={FitH},    
    pdftitle={Non-Relativistic  Four Dimensional p-Brane Supersymmetric Theories and Lie Algebra Expansion},    
    pdfauthor={Luca Romano},     
    pdfsubject={},   
    pdfnewwindow=true,      
    colorlinks=false,       
    linkcolor=red,          
    citecolor=red,        
    filecolor=red,      
    urlcolor=red           
    linkbordercolor={red},
    citebordercolor={red},
    urlbordercolor={red}
}

\usepackage{listings} 

\makeatletter

\newcommand{\Rmnum}[1]{\expandafter\@slowromancap\romannumeral #1@}
\makeatother

\theoremstyle{definition}

\theoremstyle{remark}

\theoremstyle{proposition}

\usepackage{alphalph}

\makeatletter
\newalphalph{\aalphalph}[mult]{\alphalph@alph}{26}
\newcommand{\alphalphval}[1]{%
  \@ifundefined{c@#1}{
    \aalphalph{#1}
  }{%
    \aalphalph{\value{#1}}
  }
}
\makeatother

\AtBeginEnvironment{subequations}{%
  \let\alph\alphalphval%
}

\usepackage{color}

\def\chapterautorefname~#1\null{Chap.~(#1)\null}
\def\sectionautorefname~#1\null{Sec.~(#1)\null}
\def\subsectionautorefname~#1\null{sub--Sec.~(#1)\null}
\def\figureautorefname~#1\null{Fig.~(#1)\null}
\def\tableautorefname~#1\null{Tab.~(#1)\null}
\def\equationautorefname~#1\null{eq.~(#1)\null}

\def\equationautorefname~#1\null{eq.~(#1)\null}

\NewEnviron{multieq}[1][2]{

\begin{multicols}{#1}
\begin{subequations}
\setlength{\abovedisplayskip}{-15pt}
\allowdisplaybreaks
\begin{align}
\BODY
\end{align}
\end{subequations}
\end{multicols}
\setlength{\parindent}{0pt}
}

\DeclareMathAlphabet\mathbfcal{OMS}{cmsy}{b}{n}

\usepackage[yyyymmdd]{datetime}

\title{\bf  Non-Relativistic  Four Dimensional\\ [0.3cm] p-Brane Supersymmetric Theories\\ [0.3cm]  and Lie Algebra Expansion}
\date{}

\begin{document}

\begin{flushright}
\small
IFT-UAM/CSIC-19-86\\
\normalsize
\end{flushright}
{\let\newpage\relax\maketitle}
\maketitle
\def\equationautorefname~#1\null{eq.~(#1)\null}
\def\tableautorefname~#1\null{tab.~(#1)\null}

\vspace{0.8cm}

\begin{center}


\renewcommand{\thefootnote}{\alph{footnote}}
{\sl\large Luca Romano}\footnote{Email: {\tt lucaromano2607[at]gmail.com}}

\setcounter{footnote}{0}
\renewcommand{\thefootnote}{\arabic{footnote}}

\vspace{0.5cm}

{\it Instituto de F\'{\i}sica Te\'orica UAM/CSIC\\
C/ Nicol\'as Cabrera, 13--15,  C.U.~Cantoblanco, E-28049 Madrid, Spain}\\

\vspace{1.8cm}


{\bf Abstract}
\end{center}
\begin{quotation}
  {\small We apply the Lie algebra expansion method to the $\mathcal{N}=1$ super-Poincaré algerba in four dimensions.  We define a set of p-brane projectors that induce a decomposition of the super-Poincaré algebra preparatory for the expansion. We show that starting from the $\mathcal{N}=1$ supergravity action in four dimensions it is possible to obtain two non-relativistic supersymmetric theories, one describing strings, the other membranes.}
\end{quotation}

\newpage
\pagestyle{plain}

\tableofcontents

\section*{Introduction}\addcontentsline{toc}{section}{Introduction}

In recent years non-relativistic gravity theories have received a growing interest for their role in a wide range of physical contexts, as condensed matter physics, where gravity is used as background to build effective theories, and holography, where, as an example, Newton-Cartan geometry has been recognized as the boundary geometry of a certain class of z=2 Lifshitz
spacetime \cite{Hoyos:2011ez, Son:2013rqa, Geracie:2014nka, Christensen:2013lma, Christensen:2013rfa}.\\

Among the wide class of non-relativistic gravity theories a prominent role is covered by Newtonian gravity and its geometrical re-formulation, Newton-Cartan gravity \cite{ASENS_1924_3_41__1_0, ASENS_1923_3_40__325_0}. This theory has required a specific geometrical setting, Newton-Cartan geometry, that plays, for the Newtonian gravity the same role played by Riemannian geometry for General Relativity \cite{Dixon1975,Kunzle1976, Friedrichs1928, Dautcourt1964, Trautmann1963, Duval:1984cj, RevModPhys.36.938, 1965gere.book....1T,Duval:2014uoa, Duval:2009vt}.\\

In \cite{Andringa:2010it} it has been show how starting from the Bargmann algebra , the central extension of the Galilei algebra, through a gauging procedure, it is possible to obtain Newton-Cartan formulation of Newtonian gravity. This procedure has been generalized to the stringy case and extension of the Galilei algebra \cite{Andringa:2012uz, Ozdemir:2019orp, Hansen:2018ofj}.\\ 

Recently this gauging procedure has been combined with the Lie algebra expansion method  \cite{Hatsuda:2001pp, deAzcarraga:2007et, deAzcarraga:2011pa, deAzcarraga:2012zv, Izaurieta:2006zz, deAzcarraga:2002xi} to reproduce in a systematic way the results of \cite{Andringa:2012uz, Ozdemir:2019orp, Hansen:2018ofj} and to define a method to build non-relativistic actions \cite{Bergshoeff:2019ctr}. In particular it is possible, starting from the Poincaré group, and expanding it opportunely, to generate an extension of the Galilei algebra. The expansion induced on the gauge fields defines the key to obtain non-relativistic actions from an initial relativistic action.\\

Along the the same way as for purely bosonic theories a raising attention has been direct to supersymmetric extension of non-relativistic theories. In particular, promoting the Bargmann algebra to a superalgebra, and refining the gauge procedure with opportune curvature constraints, this approach has led to the description of three dimensional non-relativistic  supersymmetric theories  \cite{Bergshoeff:2016lwr, Bergshoeff:2015ija, Andringa_2013}. In these cases the Galilei group of the bosonic case is replaced by the super-Galilei group. This suggests  a natural way to extend the systematic procedure of \cite{Bergshoeff:2019ctr}. In particular it is possible to start from the super-Poincaré algebra and use the Lie algebra expansion method to obtain the super-Galilei algebra and its extended versions.  Several results have been obtained in this direction, via the Lie algebra expansion \cite{deAzcarraga:2019mdn}, or the semigroup expansion method \cite{Concha:2014xfa, Concha:2018jxx}. In particular in \cite{deAzcarraga:2019mdn}  the Lie algebra expansion has been applied to the $\mathcal{N}=2$ super-Poincaré algebra in three dimensions and the Chern-Simons action.\\

In the present paper we extend the results of \cite{deAzcarraga:2019mdn}  to the $\mathcal{N}=1$ four dimensional super-Poincaré algebra discussing string and membrane cases. We will see that both for strings and membranes it is possible to define a non-relativistic action. Furthermore our discussion will define a general setting that could be applied to obtain different $D$-dimensional non-relativistic or ultra-relativistic $p$-brane supergravity theories. \\

The paper is organized as follows. In \autoref{sec:generalsetting} we describe the general setting; starting from the super-Poincaré algebra we define an opportune index splitting and derive the fundamental ingredients for the gauging procedure. In \autoref{sec:N=1 D=4} we specify our analysis to the four dimensional $\mathcal{N}=1$ super-Poincaré algebra and we present our main results. We study the $D=4$ $\mathcal{N}=1$ supergravity action in 1.5 order formalism. Through the application of the Lie algebra expansion method to the four dimensional super-Poincaré algebra we derive two non-relativistic actions, one describing strings, the other describing membranes. In \autoref{appendix:susyalgb} we study the technical details that are fundamental to define the Lie algebra expansion. In particular we define a set of $p$-brane projectors that provide us with a super-Poincaré algebra decomposition opportunely tailored to obtain a non-relativistic $p$-brane algebras after the expansion. Our discussions try to be as general as possible and could be immediately applied to higher dimensional cases and $\mathcal{N}>1$ as well.

\section{Lie Algebra Expansion: the Super-Poincaré Algebra }\label{sec:generalsetting}

In this section we discuss the $\mathcal{N}=1$ super-Poincaré algebra and prepare the setting for the analysis next to come. We discuss the algebra on a general ground that could be immediately applied to any dimension. We introduce the associated gauge fields and parameters and define a decomposition, using the results of \autoref{appendix:susyalgb}, and the associated Lie superalgebra expansion. For the details about the Lie algebra expansion method we refer to \cite{deAzcarraga:2002xi, Bergshoeff:2017dqq}.

\subsection{General Setting}\label{sec:generalcase}

We are going to define our setup for the analysis of the $\mathcal{N}=1$ super-Poincaré algebra in four dimensions and its expansion.  We discuss our setting with a general approach, assuming the superalgebra we are going to study is $D$-dimensional in a space with signature $(+,\ -,\ ...,\ -)$. Although the algebra we are going to study could require some case by case modifications for  $D\neq 4$, for our purpose it is convenient to proceed as described, since our analysis lends itself to immediate generalizations.\\
 
We consider the following commutation relations for our $D$-dimensional $\mathcal{N}=1$ super-Poincaré algebra \cite{ortin_2004}
\begin{subequations}
\begin{align}
[P_{\hat{A}},P_{\hat{B}}]&=0\\
[J_{\hat{A}\hat{B}},P_{\hat{C}}]&=2\eta_{\hat{C}[\hat{B}}P_{\hat{A}]}\\
[J_{\hat{A}\hat{B}},J_{\hat{C}\hat{D}}]&=
4\eta_{[\hat{A}[\hat{C}}J_{\hat{D}]\hat{B}]},\\
\{Q^{\alpha},Q^{\beta}\}&=i(\gamma^{\hat{A}}C^{-1})^{\alpha\beta}P_{\hat{A}}\\
[J_{\hat{A}\hat{B}},Q^{\alpha}]&=-\frac{1}{2}\left(\gamma_{\hat{A}\hat{B}}\right)^{\alpha}{}_{\beta}Q^{\beta}\ ,
\end{align}\label{eq:superPoincareAlgebraCommutatorsGeneralForm}
\end{subequations}
where the hatted indices run over $\hat{A}=0,...,D-1$ and the $Q$ are Majorana spinors.\footnote{Note with this choice the generators $J$ are antiHermitian and $\Omega_{\mu}^{\hat{A}\hat{B}}$ is real.}\\
Defining gauge fields and parameters as
\begin{multieq}[3]
J_{\hat{A}\hat{B}}&\rightarrow\ \Omega_{\mu}^{\hat{A}\hat{B}}\\
J_{\hat{A}\hat{B}}&\rightarrow\ \lambda^{\hat{A}\hat{B}}\\
P_{\hat{A}}&\rightarrow\ E_{\mu}^{\hat{A}}\\
P_{\hat{A}}&\rightarrow\ \eta^{\hat{A}}\\
Q^{\alpha}&\rightarrow\ \overline{\psi}_{\mu \alpha}\\
Q^{\alpha}&\rightarrow\ \overline{\epsilon}_{\alpha}
\end{multieq}
we write explicitly the gauge curvatures
\begin{subequations}
\begin{align}
R^{\hat{A}\hat{B}}(J)&=d\Omega^{\hat{A}\hat{B}}-\Omega^{\hat{A}\hat{C}}\wedge \Omega^{\hat{B}}{}_{\hat{C}}\\
R^{\hat{A}}(P)&=dE^{\hat{A}}+\Omega^{\hat{A}\hat{B}}\wedge E_{\hat{B}}-\frac{i}{2}\overline{\psi}\wedge \gamma^{\hat{A}}\psi\\
R^{\alpha}(Q)&=d\psi^{\alpha}+\frac{1}{4}\Omega^{\hat{A}\hat{B}}\wedge(\gamma_{\hat{A}\hat{B}}\psi)^{\alpha},
\end{align}
\end{subequations}
the transformations laws for the gauge fields
\begin{subequations}
\begin{align}
\delta \Omega^{\hat{A}\hat{B}}&=d\lambda^{\hat{A}\hat{B}}-2\Omega^{[\hat{A}}{}_{\hat{C}}\lambda^{\hat{B}]\hat{C}}
\\
\delta E^{\hat{A}}&=d\eta^{\hat{A}}-\lambda^{\hat{A}\hat{B}}E_{\hat{B}}+\eta_{\hat{B}}\Omega^{\hat{A}\hat{B}}+i\overline{\epsilon}\gamma^{\hat{A}}\psi\\
\delta \psi^{\alpha}&=\left[\left(d+\frac{1}{4}\Omega^{\hat{A}\hat{B}}\gamma_{\hat{A}\hat{B}}\right)\epsilon\right]^{\alpha}-\frac{1}{4}\lambda^{\hat{A}\hat{B}}(\gamma_{\hat{A}\hat{B}}\psi)^{\alpha}
\end{align}
\end{subequations}
and for the associated curvatures
\begin{subequations}
\begin{align}
\delta R^{\hat{A}\hat{B}}(J)&=-\lambda^{[\hat{B}}{}_{\hat{C}}R^{\hat{A}]\hat{C}}(J)\\
\delta R^{\hat{A}}(P)&=-\lambda^{\hat{A}\hat{B}}R_{\hat{B}}(P)+\eta_{\hat{B}}R^{\hat{A}\hat{B}}(J)+i\overline{\epsilon}\gamma^{\hat{A}}R(Q)\\
\delta R^{\alpha}(Q)&=-\frac{1}{4}\lambda^{\hat{A}\hat{B}}(\gamma_{\hat{A}\hat{B}}R(Q))^{\alpha}+\frac{1}{4}R^{\hat{A}\hat{B}}(J)(\gamma_{\hat{A}\hat{B}}\epsilon)^{\alpha}.
\end{align}
\end{subequations}
Now we consider a decomposition of the indices as $\hat{A}=\{A,a\}$ with $A=0,...,p$ and $a=p+1,...,D-1$. This induces a decomposition of the generators as
\begin{subequations}
\begin{align}
J_{\hat{A}\hat{B}}&\rightarrow\{J_{AB}, G_{Aa},J_{ab}\}\\
 P_{\hat{A}}&\rightarrow\{H_{A}, P_{a}\}.
\end{align}
\end{subequations}
Furthermore performing the manipulations described in \autoref{appendix:susyalgb}, using the projectors $\Pi_{\pm}^{U}$ (we always assume $\gamma_{0}$ to be in $U$, see \autoref{appendix:susyalgb} ),  we find , directly from \autoref{eq:superPoincareAlgebraCommutatorsGeneralForm}, the following commutation rules, 
\begin{subequations}
\begin{multicols}{2}
\setlength{\abovedisplayskip}{-15pt}
\allowdisplaybreaks
\begin{align}
[J_{AB},J_{CD}]&=4\eta_{[A[C}J_{D]B]}\\
[J_{ab},J_{cd}]&=4\eta_{[a[c}J_{d]b]}\\
[J_{AB},G_{Cd}]&=2\eta_{C[B}G_{A]d}\\
[J_{ab},G_{Cd}]&=2\eta_{d[b|}G_{C|a]}\\
[G_{Aa},G_{Bb}]&=-\eta_{AB}J_{ab}-\eta_{ab}J_{AB}\\
[J_{AB},H_{C}]&=2\eta_{C[B}H_{A]}\\
[J_{ab},P_{d}]&=2\eta_{d[b}P_{a]}\\
[G_{Aa},H_{B}]&=-\eta_{AB}P_{a}\\
[G_{Aa},P_{b}]&=\eta_{ab}H_{A}\\
[J_{AB},Q^{\alpha\pm}]&=-\frac{1}{2}\left(\gamma_{AB}\right)^{\alpha}{}_{\beta}Q^{\beta\pm}\\
[J_{ab},Q^{\alpha}]&=-\frac{1}{2}\left(\gamma_{ab}\right)^{\alpha}{}_{\beta}Q^{\beta\pm}\\
[G_{Ab},Q^{\alpha\pm}]&=-\frac{1}{2}\left(\gamma_{Ab}\right)^{\alpha}{}_{\beta}Q^{\beta\mp}\\
\{Q^{\alpha\pm},Q^{\beta\pm\xi^{U}}\}&=i\left[\Pi^{V}_{\pm}\gamma_{A}C^{-1}\right]^{\alpha\beta}H^{A}\\
\{Q^{\alpha\pm},Q^{\beta\mp \xi^{U}}\}&=i\left[\Pi^{V}_{\pm}\gamma_{a}C^{-1}\right]^{\alpha\beta}P^{a}\ ,
\end{align}\label{eq:superPoincareAlgebraCommutators}
\end{multicols}
\end{subequations}
where $\xi^{U}=(-1)^{p(p+1)/2+1}$ and the matrix $C$ appearing in the algebra is the charge conjugation matrix that in four dimensions acts as
\begin{multieq}[3]
C\gamma^{\hat{A}}C^{-1}&=-(\gamma^{\hat{A}})^{t}\\
C^{t}&=-C\\
C^{2}=-1.
\end{multieq}
We introduce the gauge fields defined by the index splitting
\begin{multicols}{3}
\begin{subequations}
\setlength{\abovedisplayskip}{-15pt}
\allowdisplaybreaks
\begin{align}
J_{AB}&\rightarrow\ \Omega_{\mu}^{AB}\\
J_{ab}&\rightarrow\ \Omega_{\mu}^{ab}\\
G_{Ab}&\rightarrow\ \Omega_{\mu}^{Ab}\\
P_{a}&\rightarrow\ E_{\mu}^{a}\\
H_{A}&\rightarrow\ \tau_{\mu}^{A}\\
Q^{\alpha +}&\rightarrow\ \overline{\psi}_{\mu\alpha}^{+}\\
Q^{\alpha -}&\rightarrow\ \overline{\psi}_{\mu\alpha}^{-}
\end{align}
\end{subequations}
\end{multicols}
\setlength{\parindent}{0pt}
The corresponding curvatures read (omitting form indices and spinor indices)
\begin{subequations}
\begin{align}
R^{AB}(J)=&d\Omega^{AB}-\Omega^{AC}\wedge \Omega^{B}_{\phantom{B}C}-\Omega^{Aa}\wedge\Omega^{B}{}_{a}\\
R^{ab}(J)=&d\Omega^{ab}-\Omega^{ac}\wedge \Omega^{b}_{\phantom{b}c}-\Omega^{Aa}\wedge\Omega_{A}^{\ b}\\
R^{Aa}(G)=&d\Omega^{Aa}+\Omega^{AB}\wedge \Omega_{B}^{\phantom{B}a}+\Omega^{ab}\wedge\Omega^{A}_{\phantom{A}b}\\
R^{a}(P)=&dE^{a}+\Omega^{ab}\wedge E_{b}-\Omega^{Aa}\wedge \tau_{A}
-\frac{i}{2}\overline{\psi}^{+}\gamma^{a}\wedge\psi^{-}-\frac{i}{2}\overline{\psi}^{-}\gamma^{a}\wedge\psi^{+}\\
R^{A}(H)=&d\tau^{A}+\Omega^{AB}\wedge \tau_{B}+\Omega^{Aa}\wedge E_{a}-\frac{i}{2}\overline{\psi}^{-}\gamma^{A}\wedge\psi^{-}-\frac{i}{2}\overline{\psi}^{+}\gamma^{A}\wedge\psi^{+}\\
R^{+}(Q)&=d\psi^{+}+\frac{1}{4}\Omega^{AB}\gamma_{AB}\wedge\psi^{+}+
\frac{1}{4}\Omega^{ab}\gamma_{ab}\wedge\psi^{+}+\frac{1}{2}\Omega^{Ab}\gamma_{Ab}\wedge\psi^{-}\\
R^{-}(Q)&=d\psi^{-}+\frac{1}{4}\Omega^{AB}\gamma_{AB}\wedge\psi^{-}+
\frac{1}{4}\Omega^{ab}\gamma_{ab}\wedge\psi^{-}+\frac{1}{2}\Omega^{Ab}\gamma_{Ab}\wedge\psi^{+}.
\end{align}
\end{subequations}
Taking the following gauge parameters
\begin{multicols}{3}
\begin{subequations}
\setlength{\abovedisplayskip}{-15pt}
\allowdisplaybreaks
\begin{align}
J_{AB}&\rightarrow\ \lambda^{AB}\\
J_{ab}&\rightarrow\ \lambda^{ab}\\
G_{Ab}&\rightarrow\ \lambda^{Ab}\\
P_{a}&\rightarrow\ \eta^{a}\\
H_{A}&\rightarrow\ \eta^{A}\\
Q^{\alpha +}&\rightarrow\ \overline{\epsilon}_{\alpha}^{+}\\
Q^{\alpha -}&\rightarrow\ \overline{\epsilon}_{\alpha}^{-}
\end{align}
\end{subequations}
\end{multicols}
\setlength{\parindent}{0pt}
we can write the transformation rules as
\begin{subequations}
\begin{align}
\delta \Omega^{AB}&= d\lambda^{AB}+\lambda^{C[A}\Omega_{C}{}^{B]}+\lambda^{a[A}\Omega_{a}{}^{B]}\\
\delta \Omega^{ab}&= d\lambda^{ab}+\lambda^{d[a}\Omega_{d}{}^{b]}+\lambda^{A[a}\Omega_{A}{}^{b]}\\
\delta \Omega^{Ab}&= d\lambda^{Ab}+\lambda^{d[A}\Omega_{d}{}^{a]}+\lambda^{B[A}\Omega_{B}{}^{a]}\\
\delta E^{a}&=d\eta^{a}+\eta_{B}\Omega^{aB}+\eta_{b}\Omega^{ab}-\lambda^{aB}\tau_{B}-\lambda^{ab}E_{b}+i\overline{\epsilon}^{+}\gamma^{a}\wedge\psi^{-}+i\overline{\epsilon}^{-}\gamma^{a}\wedge\psi^{+}\\
\delta \tau^{A}&=d\eta^{A}+\eta_{B}\Omega^{AB}+\eta_{b}\Omega^{Ab}-\lambda^{AB}\tau_{B}-\lambda^{Ab}E_{b}+i\overline{\epsilon}^{+}\gamma^{A}\wedge\psi^{+}+i\overline{\epsilon}^{-}\gamma^{A}\wedge\psi^{-}\\
\delta \psi^{+}&=d\epsilon^{+}-\frac{1}{4}\lambda^{AB}\gamma_{AB}\psi^{+}-\frac{1}{4}\Omega^{AB}\gamma_{AB}\epsilon^{+}-\frac{1}{2}\lambda^{ab}\gamma_{ab}\psi^{+}-\frac{1}{4}\Omega^{ab}\gamma_{ab}\epsilon^{+}-\frac{1}{2}\lambda^{Ab}\gamma_{Ab}\psi^{-}-\frac{1}{2}\Omega^{Ab}\gamma_{Ab}\epsilon^{-}\\
\delta \psi^{-}&=d\epsilon^{-}-\frac{1}{4}\lambda^{AB}\gamma_{AB}\psi^{-}-\frac{1}{4}\Omega^{AB}\gamma_{AB}\epsilon^{-}-\frac{1}{2}\lambda^{ab}\gamma_{ab}\psi^{-}-\frac{1}{4}\Omega^{ab}\gamma_{ab}\epsilon^{-}-\frac{1}{2}\lambda^{Ab}\gamma_{Ab}\psi^{+}-\frac{1}{2}\Omega^{Ab}\gamma_{Ab}\epsilon^{+}.
\end{align}
\end{subequations}
We recall that 
\begin{align}
\overline{\chi}^{\pm}=(\chi^{\pm \xi^{U}})^{t}C.
\end{align}
for any spinor $\chi$, see \autoref{sec:Majorana}. We have defined the general setting for the analysis of the Lie algebra expansion and the specialization to four dimensions. 

\subsection{Expansion}
In order to define a Lie superalgebra expansion we need to decompose our starting super-Poincaré algebra $\mathfrak{g}$ in two subspaces such that they have a symmetric space structure, $\mathfrak{g}=V_{0}\oplus V_{1}$. Furthermore the choice of the subalgebra $V_{0}$ is dictated by the fact that the expanded algebra $\mathfrak{g}(0,1)$ will be the Inonu-Wigner contraction of $\mathfrak{g}$ with respect to $V_{0}$ (see \cite{deAzcarraga:2002xi}). Since we are interested in a non-relativistic algebra we would choose $V_{0}$ in such a way that $\mathfrak{g}(0,1)$ corresponds to the super-Galilei algebra. As  discussed in \autoref{appendix:susyalgb} the second of these requests could be satisfied if $\xi^{U}=+1$, that corresponds to the cases $p=1$ and $p=2$. Using these information and the setting defined in the previous subsection we can choose the decomposition as
\begin{subequations}
\begin{align}
V_{0}&=\{J_{AB},J_{ab}, H_{A}, Q^{\alpha +}\}\\
V_{1}&=\{G_{Ab},P_{a}, Q^{\alpha -}\}
\end{align}
\end{subequations}

We recognize a symmetric space structure
\begin{multieq}[3]
[V_{0},V_{0}]&\subset V_{0}\\
[V_{0},V_{1}]&\subset V_{1}\\
[V_{1},V_{1}]&\subset V_{0},
\end{multieq}
with $[\ ,\ ]$ now denoting supercommutator 
\begin{align}
[X,Y]=XY-(-1)^{\deg X \deg Y}YX,
\end{align}
where $\deg X$ is 0 or 1 if $X$ belongs respectively to $V_{0}$ or $V_{1}$. This results in a commutator or anticommutator depending on the nature of the arguments. This allows us to apply theorem 2 and proposition 2 of \cite{deAzcarraga:2002xi} and expand the gauge fields as follows 
\begin{subequations}
\begin{alignat}{2}
\Omega_{\mu}^{AB}=&\sum_{\substack{k=0 \\ k\in 2\mathbb{Z}_{+}}}^{N_{0}}\lambda^{k}\ \accentset{(k)}{\Omega}_{\mu}^{AB}&&=\accentset{(0)}{\Omega}_{\mu}^{AB}+\lambda^2\ \accentset{(2)}{\Omega}_{\mu}^{AB}+...\\
\Omega_{\mu}^{ab}=&\sum_{\substack{k=0 \\ k\in 2\mathbb{Z}_{+}}}^{N_{0}}\lambda^{k}\ \accentset{(k)}{\Omega}_{\mu}^{ab}&&=\accentset{(0)}{\Omega}_{\mu}^{ab}+\lambda^2\ \accentset{(2)}{\Omega}_{\mu}^{ab}+...\\
\tau^{A}_{\mu}=&\sum_{\substack{k=0 \\ k\in 2\mathbb{Z}_{+}}}^{N_{0}}\lambda^{k}\ \accentset{(k)}{\tau}_{\mu}^{A}&&=\ \accentset{(0)}{\tau}_{\mu}^{A}+\lambda^{2}\ \accentset{(2)}{\tau}_{\mu}^{A}+...\\
\psi^{+}_{\alpha}=&\sum_{\substack{k=0 \\ k\in 2\mathbb{Z}_{+}}}^{N_{0}}\lambda^{k}\ \accentset{(k)}{\psi}^{+}_{\alpha}&&=\ \accentset{(0)}{\psi}^{+}_{\alpha}+\lambda^{2}\ \accentset{(2)}{\psi}^{+}_{\alpha}+...\\
E_{\mu}^{a}=&\sum_{\substack{k=1 \\ k\in 2\mathbb{Z}_{+}+1}}^{N_{1}}\lambda^{k}\ \accentset{(k)}{E}_{\mu}^{a}&&=\lambda\ \accentset{(1)}{E}_{\mu}^{a}+\lambda^3\ \accentset{(3)}{E}_{\mu}^{a}+...\\
\Omega_{\mu}^{Ab}=&\sum_{\substack{k=1 \\ k\in 2\mathbb{Z}_{+}+1}}^{N_{1}}\lambda^{k}\ \accentset{(k)}{\Omega}_{\mu}^{Ab}
&&=\lambda \ \accentset{(1)}{\Omega}_{\mu}^{Ab}+\lambda^3 \ \accentset{(3)}{\Omega}_{\mu}^{Ab}+...\\
\psi^{-}_{\alpha}=&\sum_{\substack{k=1 \\ k\in 2\mathbb{Z}_{+}+1}}^{N_{1}}\lambda^{k}\ \accentset{(k)}{\psi}^{-}_{\alpha}&&=\lambda\ \accentset{(1)}{\psi}^{-}_{\alpha}+\lambda^3\ \accentset{(3)}{\psi}^{-}_{\alpha}+...\ , \\
\end{alignat}
\end{subequations}
where $N_{1}$ is odd and $N_{0}$ is even.  A consistent truncation requires $N_{1}=N_{0}\pm 1$. The algebra obtained by this way is denoted with $\mathfrak{g}(N_{0},N_{1})$ \cite{deAzcarraga:2002xi}.

\subsubsection{An Example: $\mathbf{\mathfrak{g}(4,3)}$}
As an example let's inspect the commutation relations obtained by applying the Lie algebra expansion method  of \cite{deAzcarraga:2002xi} to the algebra \autoref{eq:superPoincareAlgebraCommutators} (with $\xi^{U}=+1$) up to the order $N_{0}=4,\ N_{1}=3$. We obtain for the non-zero commutators
\begin{multicols}{2}
\begin{subequations}
\setlength{\abovedisplayskip}{-15pt}
\allowdisplaybreaks
\begin{align}
[\accentset{(0)}{J}_{AB},\accentset{(0)}{J}_{CD}]&=4\eta_{[A[C}\accentset{(0)}{J}_{D]B]}\\
[\accentset{(0)}{J}_{AB},\accentset{(2)}{J}_{CD}]&=4\eta_{[A[C}\accentset{(2)}{J}_{D]B]}\\
[\accentset{(0)}{J}_{AB},\accentset{(4)}{J}_{CD}]&=4\eta_{[A[C}\accentset{(4)}{J}_{D]B]}\\
[\accentset{(0)}{J}_{ab},\accentset{(0)}{J}_{cd}]&=4\eta_{[a[c}\accentset{(0)}{J}_{d]b]}\\
[\accentset{(0)}{J}_{ab},\accentset{(2)}{J}_{cd}]&=4\eta_{[a[c}\accentset{(2)}{J}_{d]b]}\\
[\accentset{(0)}{J}_{ab},\accentset{(4)}{J}_{cd}]&=4\eta_{[a[c}\accentset{(4)}{J}_{d]b]}\\
[\accentset{(0)}{J}_{AB},\accentset{(1)}{G}_{Cd}]&=2\eta_{C[B}\accentset{(1)}{G}_{A]d}\\
[\accentset{(0)}{J}_{AB},\accentset{(3)}{G}_{Cd}]&=2\eta_{C[B}\accentset{(3)}{G}_{A]d}\\
[\accentset{(2)}{J}_{AB},\accentset{(1)}{G}_{Cd}]&=2\eta_{C[B}\accentset{(3)}{G}_{A]d}\\
[\accentset{(0)}{J}_{ab},\accentset{(1)}{G}_{Cd}]&=2\eta_{d[b|}\accentset{(1)}{G}_{C|a]}\\
[\accentset{(0)}{J}_{ab},\accentset{(3)}{G}_{Cd}]&=2\eta_{d[b|}\accentset{(3)}{G}_{C|a]}\\
[\accentset{(2)}{J}_{ab},\accentset{(1)}{G}_{Cd}]&=2\eta_{d[b|}\accentset{(3)}{G}_{C|a]}\\
[\accentset{(1)}{G}_{Aa},\accentset{(1)}{G}_{Bb}]&=-\eta_{AB}\accentset{(2)}{J}_{ab}-\eta_{ab}\accentset{(2)}{J}_{AB}\\
[\accentset{(1)}{G}_{Aa},\accentset{(3)}{G}_{Bb}]&=-\eta_{AB}\accentset{(4)}{J}_{ab}-\eta_{ab}\accentset{(4)}{J}_{AB}\\
[\accentset{(0)}{J}_{AB},\accentset{(0)}{H}_{C}]&=2\eta_{C[B}\accentset{(0)}{H}_{A]}\\
[\accentset{(0)}{J}_{AB},\accentset{(2)}{H}_{C}]&=2\eta_{C[B}\accentset{(2)}{H}_{A]}\\
[\accentset{(2)}{J}_{AB},\accentset{(0)}{H}_{C}]&=2\eta_{C[B}\accentset{(2)}{H}_{A]}\\
[\accentset{(0)}{J}_{AB},\accentset{(4)}{H}_{C}]&=2\eta_{C[B}\accentset{(4)}{H}_{A]}\\
[\accentset{(4)}{J}_{AB},\accentset{(0)}{H}_{C}]&=2\eta_{C[B}\accentset{(4)}{H}_{A]}\\
[\accentset{(2)}{J}_{AB},\accentset{(2)}{H}_{C}]&=2\eta_{C[B}\accentset{(4)}{H}_{A]}\\
[\accentset{(0)}{J}_{ab},\accentset{(1)}{P}_{d}]&=2\eta_{d[b}\accentset{(1)}{P}_{a]}\\
[\accentset{(0)}{J}_{ab},\accentset{(3)}{P}_{d}]&=2\eta_{d[b}\accentset{(3)}{P}_{a]}\\
[\accentset{(2)}{J}_{ab},\accentset{(1)}{P}_{d}]&=2\eta_{d[b}\accentset{(3)}{P}_{a]}\\
[\accentset{(1)}{G}_{Aa},\accentset{(0)}{H}_{B}]&=-\eta_{AB}\accentset{(1)}{P}_{a}\\
[\accentset{(1)}{G}_{Aa},\accentset{(2)}{H}_{B}]&=-\eta_{AB}\accentset{(3)}{P}_{a}\\
[\accentset{(3)}{G}_{Aa},\accentset{(0)}{H}_{B}]&=-\eta_{AB}\accentset{(3)}{P}_{a}\\
[\accentset{(1)}{G}_{Aa},\accentset{(1)}{P}_{b}]&=\eta_{ab}\accentset{(2)}{H}_{A}\\
[\accentset{(1)}{G}_{Aa},\accentset{(3)}{P}_{b}]&=\eta_{ab}\accentset{(4)}{H}_{A}\\
[\accentset{(3)}{G}_{Aa},\accentset{(1)}{P}_{b}]&=\eta_{ab}\accentset{(4)}{H}_{A}\\
\{\accentset{(0)}{Q}^{\alpha +},\accentset{(0)}{Q}^{\beta +}\}&=i\left[\Pi^{U}_{+}\gamma_{A}C^{-1}\right]^{\alpha\beta}\accentset{(0)}{H}^{A}\\
\{\accentset{(0)}{Q}^{\alpha +},\accentset{(2)}{Q}^{\beta +}\}&=i\left[\Pi^{U}_{+}\gamma_{A}C^{-1}\right]^{\alpha\beta}\accentset{(2)}{H}^{A}\\
\{\accentset{(0)}{Q}^{\alpha +},\accentset{(4)}{Q}^{\beta +}\}&=i\left[\Pi^{U}_{+}\gamma_{A}C^{1}\right]^{\alpha\beta}\accentset{(4)}{H}^{A}\\
\{\accentset{(2)}{Q}^{\alpha +},\accentset{(2)}{Q}^{\beta +}\}&=i\left[\Pi^{U}_{+}\gamma_{A}C^{-1}\right]^{\alpha\beta}\accentset{(4)}{H}^{A}\\
\{\accentset{(1)}{Q}^{\alpha -},\accentset{(1)}{Q}^{\beta -}\}&=i\left[\Pi^{U}_{-}\gamma_{A}C^{-1}\right]^{\alpha\beta}\accentset{(2)}{H}^{A}\\
\{\accentset{(1)}{Q}^{\alpha -},\accentset{(3)}{Q}^{\beta -}\}&=i\left[\Pi^{U}_{-}\gamma_{A}C^{-1}\right]^{\alpha\beta}\accentset{}{H}^{A}\\
\{\accentset{(0)}{Q}^{\alpha +},\accentset{(1)}{Q}^{\beta -}\}&=i\left[\Pi^{U}_{+}\gamma_{a}C^{-1}\right]^{\alpha\beta}\accentset{(1)}{P}^{a}\\
\{\accentset{(0)}{Q}^{\alpha +},\accentset{(3)}{Q}^{\beta -}\}&=i\left[\Pi^{U}_{+}\gamma_{a}C^{-1}\right]^{\alpha\beta}\accentset{(3)}{P}^{a}\\
\{\accentset{(2)}{Q}^{\alpha +},\accentset{(1)}{Q}^{\beta -} \}&=i\left[\Pi^{U}_{+}\gamma_{a}C^{-1}\right]^{\alpha\beta}\accentset{(3)}{P}^{a}\\
[\accentset{(0)}{J}_{AB},\accentset{(0)}{Q}^{\alpha +}]&=-\frac{1}{2}(\gamma_{AB})^{\alpha}{}_{\beta}\accentset{(0)}{Q}^{\beta +}\\
[\accentset{(0)}{J}_{AB},\accentset{(2)}{Q}^{\alpha +}]&=-\frac{1}{2}(\gamma_{AB})^{\alpha}{}_{\beta}\accentset{(2)}{Q}^{\beta +}\\
[\accentset{(2)}{J}_{AB},\accentset{(0)}{Q}^{\alpha +}]&=-\frac{1}{2}(\gamma_{AB})^{\alpha}{}_{\beta}\accentset{(2)}{Q}^{\beta +}\\
[\accentset{(0)}{J}_{AB},\accentset{(4)}{Q}^{\alpha +}]&=-\frac{1}{2}(\gamma_{AB})^{\alpha}{}_{\beta}\accentset{(4)}{Q}^{\beta +}\\
[\accentset{(4)}{J}_{AB},\accentset{(0)}{Q}^{\alpha +}]&=-\frac{1}{2}(\gamma_{AB})^{\alpha}{}_{\beta}\accentset{(4)}{Q}^{\beta +}\\
[\accentset{(2)}{J}_{AB},\accentset{(2)}{Q}^{\alpha +}]&=-\frac{1}{2}(\gamma_{AB})^{\alpha}{}_{\beta}\accentset{(4)}{Q}^{\beta +}\\
[\accentset{(0)}{J}_{AB},\accentset{(1)}{Q}^{\alpha -}]&=-\frac{1}{2}(\gamma_{AB})^{\alpha}{}_{\beta}\accentset{(1)}{Q}^{\beta -}\\
[\accentset{(0)}{J}_{AB},\accentset{(3)}{Q}^{\alpha -}]&=-\frac{1}{2}(\gamma_{AB})^{\alpha}{}_{\beta}\accentset{(3)}{Q}^{\beta -}\\
[\accentset{(2)}{J}_{AB},\accentset{(1)}{Q}^{\alpha -}]&=-\frac{1}{2}(\gamma_{AB})^{\alpha}{}_{\beta}\accentset{(3)}{Q}^{\beta -}\\
[\accentset{(0)}{J}_{ab},\accentset{(0)}{Q}^{\alpha +}]&=-\frac{1}{2}(\gamma_{ab})^{\alpha}{}_{\beta}\accentset{(0)}{Q}^{\beta +}\\
[\accentset{(0)}{J}_{ab},\accentset{(2)}{Q}^{\alpha +}]&=-\frac{1}{2}(\gamma_{ab})^{\alpha}{}_{\beta}\accentset{(2)}{Q}^{\beta +}\\
[\accentset{(2)}{J}_{ab},\accentset{(0)}{Q}^{\alpha +}]&=-\frac{1}{2}(\gamma_{ab})^{\alpha}{}_{\beta}\accentset{(2)}{Q}^{\beta +}\\
[\accentset{(0)}{J}_{ab},\accentset{(4)}{Q}^{\alpha +}]&=-\frac{1}{2}(\gamma_{ab})^{\alpha}{}_{\beta}\accentset{(4)}{Q}^{\beta +}\\
[\accentset{(4)}{J}_{ab},\accentset{(0)}{Q}^{\alpha +}]&=-\frac{1}{2}(\gamma_{ab})^{\alpha}{}_{\beta}\accentset{(4)}{Q}^{\beta +}\\
[\accentset{(2)}{J}_{ab},\accentset{(2)}{Q}^{\alpha +}]&=-\frac{1}{2}(\gamma_{ab})^{\alpha}{}_{\beta}\accentset{(4)}{Q}^{\beta +}\\
[\accentset{(0)}{J}_{ab},\accentset{(1)}{Q}^{\alpha -}]&=-\frac{1}{2}(\gamma_{ab})^{\alpha}{}_{\beta}\accentset{(1)}{Q}^{\beta -}\\
[\accentset{(0)}{J}_{ab},\accentset{(3)}{Q}^{\alpha -}]&=-\frac{1}{2}(\gamma_{ab})^{\alpha}{}_{\beta}\accentset{(3)}{Q}^{\beta -}\\
[\accentset{(2)}{J}_{ab},\accentset{(1)}{Q}^{\alpha -}]&=-\frac{1}{2}(\gamma_{ab})^{\alpha}{}_{\beta}\accentset{(3)}{Q}^{\beta -}\\
[\accentset{(1)}{G}_{Ab},\accentset{(0)}{Q}^{\alpha +}]&=-\frac{1}{2}(\gamma_{Ab})^{\alpha}{}_{\beta}\accentset{(1)}{Q}^{\beta -}\\
[\accentset{(3)}{G}_{Ab},\accentset{(0)}{Q}^{\alpha +}]&=-\frac{1}{2}(\gamma_{Ab})^{\alpha}{}_{\beta}\accentset{(3)}{Q}^{\beta -}\\
[\accentset{(1)}{G}_{Ab},\accentset{(2)}{Q}^{\alpha +}]&=-\frac{1}{2}(\gamma_{Ab})^{\alpha}{}_{\beta}\accentset{(3)}{Q}^{\beta -}\\
[\accentset{(1)}{G}_{Ab},\accentset{(1)}{Q}^{\alpha -}]&=-\frac{1}{2}(\gamma_{Ab})^{\alpha}{}_{\beta}\accentset{(2)}{Q}^{\beta +}\\
[\accentset{(1)}{G}_{Ab},\accentset{(3)}{Q}^{\alpha -}]&=-\frac{1}{2}(\gamma_{Ab})^{\alpha}{}_{\beta}\accentset{(4)}{Q}^{\beta +}\\
[\accentset{(3)}{G}_{Ab},\accentset{(1)}{Q}^{\alpha -}]&=-\frac{1}{2}(\gamma_{Ab})^{\alpha}{}_{\beta}\accentset{(4)}{Q}^{\beta +}\ .
\end{align}\label{eq:g(4,3)}
\end{subequations}
\end{multicols}

\setlength{\parindent}{0pt}

If we consider the truncation $\mathfrak{g}(0,1)$ we find the following commutations relations
\begin{multicols}{2}
\begin{subequations}
\setlength{\abovedisplayskip}{-15pt}
\allowdisplaybreaks
\begin{align}
[\accentset{(0)}{J}_{AB},\accentset{(0)}{J}_{CD}]&=4\eta_{[A[C}\accentset{(0)}{J}_{D]B]}\\
[\accentset{(0)}{J}_{ab},\accentset{(0)}{J}_{cd}]&=4\eta_{[a[c}\accentset{(0)}{J}_{d]b]}\\
[\accentset{(0)}{J}_{AB},\accentset{(1)}{G}_{Cd}]&=2\eta_{C[B}\accentset{(1)}{G}_{A]d}\\
[\accentset{(0)}{J}_{ab},\accentset{(1)}{G}_{Cd}]&=2\eta_{d[b|}\accentset{(1)}{G}_{C|a]}\\
[\accentset{(1)}{G}_{Aa},\accentset{(1)}{G}_{Bb}]&=0\\
[\accentset{(0)}{J}_{AB},\accentset{(0)}{H}_{C}]&=2\eta_{C[B}\accentset{(0)}{H}_{A]}\\
[\accentset{(0)}{J}_{ab},\accentset{(1)}{P}_{d}]&=2\eta_{d[b}\accentset{(1)}{P}_{a]}\\
[\accentset{(1)}{G}_{Aa},\accentset{(0)}{H}_{B}]&=-\eta_{AB}\accentset{(1)}{P}_{a}\\
[\accentset{(1)}{G}_{Aa},\accentset{(1)}{P}_{b}]&=0\\
\{\accentset{(0)}{Q}^{\alpha +},\accentset{(0)}{Q}^{\beta +}\}&=i\left[\Pi^{U}_{+}\gamma_{A}C^{-1}\right]^{\alpha\beta}\accentset{(0)}{H}^{A}\\
\{\accentset{(1)}{Q}^{\alpha -},\accentset{(1)}{Q}^{\beta -}\}&=0\\
\{\accentset{(0)}{Q}^{\alpha +},\accentset{(1)}{Q}^{\beta -}\}&=i\left[\Pi^{U}_{+}\gamma_{a}C^{-1}\right]^{\alpha\beta}\accentset{(1)}{P}^{a}\\
[\accentset{(0)}{J}_{AB},\accentset{(0)}{Q}^{\alpha +}]&=-\frac{1}{2}(\gamma_{AB})^{\alpha}{}_{\beta}\accentset{(0)}{Q}^{\beta +}\\
[\accentset{(0)}{J}_{AB},\accentset{(1)}{Q}^{\alpha -}]&=-\frac{1}{2}(\gamma_{AB})^{\alpha}{}_{\beta}\accentset{(1)}{Q}^{\beta -}\\
[\accentset{(0)}{J}_{ab},\accentset{(0)}{Q}^{\alpha +}]&=-\frac{1}{2}(\gamma_{ab})^{\alpha}{}_{\beta}\accentset{(0)}{Q}^{\beta +}\\
[\accentset{(0)}{J}_{ab},\accentset{(1)}{Q}^{\alpha -}]&=-\frac{1}{2}(\gamma_{ab})^{\alpha}{}_{\beta}\accentset{(1)}{Q}^{\beta -}\\
[\accentset{(1)}{G}_{Ab},\accentset{(0)}{Q}^{\alpha +}]&=-\frac{1}{2}(\gamma_{Ab})^{\alpha}{}_{\beta}\accentset{(1)}{Q}^{\beta -}\\
[\accentset{(1)}{G}_{Ab},\accentset{(1)}{Q}^{\alpha -}]&=0\ .
\end{align}\label{eq:g(0,1)}
\end{subequations}
\end{multicols}
We recognize that this is the Inonu-Wigner contraction of the super-Poincaré algebra with respect to $V_{0}$, i.e the $p$-brane super-Galilei algebra. In particular we could identify $\mathfrak{g}(4,3)$  as an extension of the  $p$-brane super-Galilei algebra.

\section{$\mathbfcal{N}$=1 D=4 super-Poincaré Algebra}\label{sec:N=1 D=4}

In this section we focus on the $\mathcal{N}=1$ super-Poincaré algebra in four dimensions and apply the Lie algebra expansion method. We will define a non-relativistic supersymmetric action for strings and membranes.

\subsection{The Algebra}
 We report here, for convenience, the commutation relations of the $\mathcal{N}=1$ four dimensional super-Poincaré  algebra as derived in \autoref{appendix:susyalgb} for $p=1,2$ with the use of the projectors $\Pi^{U}_{\pm}$ defined in \autoref{eq:projectorsdefinition}, 
\begin{subequations}
\begin{align}
\{Q^{\alpha\pm},Q^{\beta\pm}\}&=i\left[\Pi^{U}_{\pm}\gamma_{A}C^{-1}\right]^{\alpha\beta}H^{A}\\
\{Q^{\alpha\pm},Q^{\beta\mp}\}&=i\left[\Pi^{U}_{\pm}\gamma_{a}C^{-1}\right]^{\alpha\beta}P^{a}\\
[J_{AB},Q^{\alpha\pm}]&=-\frac{1}{2}\left(\gamma_{AB}\right)^{\alpha}{}_{\beta}Q^{\beta\pm}\\
[J_{ab},Q^{\alpha\pm}]&=-\frac{1}{2}\left(\gamma_{ab}\right)^{\alpha}{}_{\beta}Q^{\beta\pm}\\
[G_{Ab},Q^{\alpha\pm}]&=-\frac{1}{2}\left(\gamma_{Ab}\right)^{\alpha}{}_{\beta}Q^{\beta\mp}\ ,
\end{align}
\end{subequations}
where $Q$ is a Majorana spinor,
\begin{align}
\overline{Q}=Q^{t}C
\end{align}
We recall that the charge conjugation matrix satisfies 
\begin{multieq}[3]
C\gamma_{\hat{A}}C^{-1}&=-\gamma^{t}_{\hat{A}}\\
C^{2}&=-1\\
C^{t}=-C\ .
\end{multieq}
We can choose the gamma matrices to be purely imaginary, an explicit choice is 
\begin{multieq}[2]
\gamma_{0}=&\sigma_{1}\otimes\sigma_{2}\\
\gamma_{1}=&i\mathds{1}\otimes\sigma_{3}\\
\gamma_{2}=&-i\sigma_{2}\otimes\sigma_{2}\\
\gamma_{3}=&-i\mathds{1}\otimes\sigma_{1}\\
\gamma_{5}=&\sigma_{3}\otimes\sigma_{2}=i\gamma_{0}\gamma_{1}\gamma_{2}\gamma_{3}\\
C=C_{(-)}=&-i\sigma_{1}\otimes\sigma_{2}=-i\gamma_{0}\\
C_{(+)}=&i\sigma_{2}\otimes\sigma_{1}=\gamma_{2}\gamma_{1}\ ,
\end{multieq}
where $\sigma_{i}$ are the Pauli matrices
\begin{multieq}[3]
\sigma_{1}&=\left(\begin{array}{cc}
0&1\\
1&0
\end{array}\right)\\
\sigma_{2}&=\left(\begin{array}{cc}
0&-i\\
i&0
\end{array}\right)\\
\sigma_{3}&=\left(\begin{array}{cc}
1&0\\
0&-1
\end{array}\right)\ .
\end{multieq}
We assume $\gamma_{0}$ to be always in $U$.  In four dimensions the projectors $\Pi^{U}_{\pm}$ satisfy the following relations 
\begin{subequations}
\begin{align}
\Pi_{\pm}^{U}\gamma_{A}&=\gamma_{A}\Pi_{\pm(-1)^{p}}^{U}\\
\Pi_{\pm}^{U}\gamma_{a}&=\gamma_{a}\Pi_{\pm(-1)^{p+1}}^{U}\\
\Pi_{\pm}^{U}\gamma_{5}&=\gamma_{5}\Pi_{\pm(-1)^{p+1}}^{U}\ .
\end{align}
\end{subequations}

Explicitly for, $p=1$ and $p=2$ we have respectively $U=-\gamma_{0}\gamma_{1}$ and $U=i\gamma_{0}\gamma_{1}\gamma_{2}$. Having defined the setting for the Lie algebra expansion of the four dimensional super-Poincaré algebra now we turn our attention to the construction of non-relativistic supersymmetric actions.

\subsection{The Action}

Our starting point to build a non-relativistic supersymmetric action in four dimension is the $\mathcal{N}=1$ $D=4$ supergravity action in 1.5 order formalism \cite{Deser:1976eh, CHAMSEDDINE197739}, namely we treat spin connection and vielbein as independent fields, but we require the spin connection to be on-shell to guarantee the invariance of the action under the supersymmetry transformations. Explicitly the Lagrangian reads  
\begin{align}
\mathcal{L}=EE^{\mu}_{\hat{A}}E^{\nu}_{\hat{B}}R^{\hat{A}\hat{B}}_{\mu\nu}(J)+2\epsilon^{\mu\nu\lambda\sigma}\overline{\psi}_{\lambda}\gamma_{5}\gamma_{\sigma} D_{\mu}\psi_{\nu}\ ,
\end{align}
that could be rewritten, up to a multiplicative constant factor, as
\begin{align}
\mathcal{L}=\epsilon_{\hat{A}\hat{B}\hat{C}\hat{D}}R^{\hat{A}\hat{B}}(J)\wedge E^{\hat{C}}\wedge E^{\hat{D}}+4\overline{\psi}\gamma_{5}\gamma_{\hat{A}}\wedge E^{\hat{A}}\wedge D\psi\ ,
\end{align}
where we have omitted the form indices. The Rarita-Schwinger term could be put in the form
\begin{align}
\overline{\psi}\gamma_{5}\gamma_{\hat{A}}\wedge E^{\hat{A}}\wedge D\psi&=\overline{\psi}\gamma_{5}\gamma_{\hat{A}}\wedge E^{\hat{A}}\wedge R(Q)\ .
\end{align}
Now we consider the string case, $p=1$, and the membrane case, $p=2$.

\subsubsection{String Case p=1}
In this case for the Rarita-Schwinger term in this case we obtain (see \autoref{appendix:susyalgb})  
\begin{align}
\overline{\psi}\gamma_{5}\gamma_{\hat{A}}\wedge E^{\hat{A}}\wedge R(Q)=&(\psi^{+})^{t}C\gamma_{5}\gamma_{A}\wedge\tau^{A}\wedge R^{+}(Q)+(\psi^{-})^{t}C\gamma_{5}\gamma_{A}\wedge\tau^{A}\wedge R^{-}(Q)+\nonumber\\
&+(\psi^{+})^{t}C\gamma_{5}\gamma_{a}\wedge E^{a}\wedge R^{-}(Q)+(\psi^{-})^{t}C\gamma_{5}\gamma_{a}\wedge E^{a}\wedge R^{+}(Q)\ .
\end{align}
Then for the full Lagrangian we can write
\begin{align}
\mathcal{L}&=\epsilon_{ABab}\bigg[R^{ab}(J)\wedge \tau^{A} \wedge \tau^{B}+2R^{Aa}(G)\wedge E^{b} \wedge \tau^{B}+ R^{AB}(J)\wedge E^{a} \wedge E^{b}
\bigg]+\nonumber\\
&+4\bigg[(\psi^{+})^{t}C\gamma_{5}\gamma_{A}\wedge\tau^{A}\wedge R^{+}(Q)+(\psi^{-})^{t}C\gamma_{5}\gamma_{A}\wedge\tau^{A}\wedge R^{-}(Q)+\nonumber\\
&+(\psi^{+})^{t}C\gamma_{5}\gamma_{a}\wedge E^{a}\wedge R^{-}(Q)+(\psi^{-})^{t}C\gamma_{5}\gamma_{a}\wedge E^{a}\wedge R^{+}(Q)\bigg]\ .
\end{align}
We can expand this Lagrangian by taking the algebra $\mathfrak{g}(2,3)$, namely imposing the truncation $N_{0}=2$ and $N_{3}=3$. We have at the lowest orders
\begin{subequations}
\begin{align}
\accentset{(0)}{\mathcal{L}}&=\epsilon_{ABab}\accentset{(0)}{R}^{AB}(J)\wedge\accentset{(0)}{E}^{a}\wedge\accentset{(0)}{E}^{b}-4\accentset{(0)}{\tau}^{A}\wedge(\accentset{(0)}{\psi}^{+})^{t}C\gamma_{5}\gamma_{A}\accentset{(0)}{R}^{+}(Q)\\
\accentset{(2)}{\mathcal{L}}&=\epsilon_{ABab}\bigg[\accentset{(0)}{R}^{AB}(J)\wedge \accentset{(1)}{E}^{a}\wedge \accentset{(1)}{E}^{b}+\accentset{(1)}{R}^{Aa}(G)\wedge \accentset{(1)}{E}^{b}\wedge \accentset{(0)}{\tau}^{B}+\accentset{(2)}{R}^{ab}(J)\wedge \accentset{(0)}{\tau}^{A}\wedge \accentset{(0)}{\tau}^{B}+2\accentset{(0)}{R}^{ab}(J)\wedge \accentset{(2)}{\tau}^{A}\wedge \accentset{(0)}{\tau}^{B}\bigg]+\nonumber\\
&-4\bigg[\accentset{(0)}{\tau}^{A}\wedge (\accentset{(2)}{\psi}^{+})^{t}C\gamma_{5}\gamma_{A}\wedge\accentset{(0)}{R}^{+}(Q)+\accentset{(0)}{\tau}^{A}\wedge (\accentset{(0)}{\psi}^{+})^{t}C\gamma_{5}\gamma_{A}\wedge\accentset{(2)}{R}^{+}(Q)+\nonumber\\
&+\accentset{(2)}{\tau}^{A}\wedge (\accentset{(0)}{\psi}^{+})^{t}C\gamma_{5}\gamma_{A}\wedge\accentset{(0)}{R}^{+}(Q)+\accentset{(0)}{\tau}^{A}\wedge (\accentset{(1)}{\psi}^{-})^{t}C\gamma_{5}\gamma_{A}\wedge\accentset{(1)}{R}^{-}(Q)+\nonumber\\
&+\accentset{(1)}{E}^{a}\wedge (\accentset{(0)}{\psi}^{+})^{t}C\gamma_{5}\gamma_{a}\wedge\accentset{(1)}{R}^{-}(Q)+\accentset{(1)}{E}^{a}\wedge (\accentset{(1)}{\psi}^{-})^{t}C\gamma_{5}\gamma_{a}\wedge\accentset{(0)}{R}^{+}(Q)\bigg]\ .
\end{align}
\end{subequations}
Among the terms discussed above our candidate action is given by taking the Lagrangian density $\accentset{(2)}{\mathcal{L}}$ . Our choice is dictated by the criteria described in \cite{Bergshoeff:2019ctr}. In particular in $\accentset{(2)}{\mathcal{L}}$ all the fields of the theory, i.e of the algebra $\mathfrak{g}(2,3)$,  appear and furthermore the truncation to $\mathfrak{g}(2,3)$ does not affect their form if compared with the untrucated algebra. These two properties guarantee that the expanded action remains invariant under the expanded transformations corresponding to the invariance of the initial action. In our case however we recognize that in the action $\accentset{(2)}{\mathcal{L}}$ the fields $\accentset{(3)}{\Omega}^{Ab}$, $\accentset{(2)}{\Omega}^{AB}$, $\accentset{(3)}{\psi}^{-}$ and $\accentset{(3)}{E}^{a}$ do not appear; as explained in \cite{Bergshoeff:2019ctr} this does not affect, in the present case, the invariance under the associated gauge transformation. Indeed it could be immediately checked that under the gauge
transformations corresponding to the generators associated with the fields above,
\begin{subequations}
\begin{align}
\delta\accentset{(2)}{\Omega}^{AB}&=d\accentset{(2)}{\lambda}^{AB}+\accentset{(2)}{\lambda}^{C[A}\accentset{(0)}{\Omega}_{C}{}^{B]}\\
\delta\accentset{(3)}{\Omega}^{Ab}&=d\accentset{(3)}{\lambda}^{Ab}+\frac{1}{2}\accentset{(3)}{\lambda}^{dA}\accentset{(0)}{\Omega}_{d}{}^{b}-\frac{1}{2}\accentset{(3)}{\lambda}^{Ba}\accentset{(0)}{\Omega}_{B}{}^{A}\\
\delta\accentset{(3)}{E}^{a}&=-\accentset{(3)}{\lambda}^{aB}\accentset{(0)}{\tau}_{B}+i(\accentset{(3)}{\epsilon})^{t}C\gamma^{a}\accentset{(0)}{\psi}^{+}\\
\delta\accentset{(3)}{\psi}^{-}&=d\accentset{(3)}{\epsilon}^{-}-
\frac{1}{4}\accentset{(2)}{\lambda}^{AB}\gamma_{AB}\accentset{(1)}{\psi}^{-}-\frac{1}{4}\accentset{(0)}{\Omega}^{AB}\gamma_{AB}\accentset{(3)}{\epsilon}^{-}
-\frac{1}{4}\accentset{(2)}{\lambda}^{ab}\gamma_{ab}\accentset{(1)}{\epsilon}^{-}-\frac{1}{2}\accentset{(3)}{\lambda}^{Ab}\gamma_{Ab}\accentset{(0)}{\psi}^{+}\\
\delta\accentset{(2)}{\tau}^{A}&=-\accentset{(2)}{\lambda}^{AB}\accentset{(0)}{\tau}_{B}\\
\delta\accentset{(2)}{\psi}^{+}&=-\frac{1}{4}\accentset{(2)}{\lambda}^{AB}\gamma_{AB}\accentset{(0)}{\psi}^{+}\ ,
\end{align}
\end{subequations}
the Einstein-Hilbert term and the Rarita-Schwinger are separately invariant (note that only the last two fields above appear in the action). Thus we have obtained the following non-relativistic supersymmetric action for string super-Galilei algebra in four dimensions
\begin{align}
\mathcal{L}^{p=1}=\epsilon_{ABab}&\bigg[\accentset{(0)}{R}^{AB}(J)\wedge \accentset{(1)}{E}^{a}\wedge \accentset{(1)}{E}^{b}+\accentset{(1)}{R}^{Aa}(G)\wedge \accentset{(1)}{E}^{b}\wedge \accentset{(0)}{\tau}^{B}+\accentset{(2)}{R}^{ab}(J)\wedge \accentset{(0)}{\tau}^{A}\wedge \accentset{(0)}{\tau}^{B}+2\accentset{(0)}{R}^{ab}(J)\wedge \accentset{(2)}{\tau}^{A}\wedge \accentset{(0)}{\tau}^{B}\bigg]+\nonumber\\
-4&\bigg[\accentset{(0)}{\tau}^{A}\wedge (\accentset{(2)}{\psi}^{+})^{t}C\gamma_{5}\gamma_{A}\wedge\accentset{(0)}{R}^{+}(Q)+\accentset{(0)}{\tau}^{A}\wedge (\accentset{(0)}{\psi}^{+})^{t}C\gamma_{5}\gamma_{A}\wedge\accentset{(2)}{R}^{+}(Q)+\nonumber\\
&+\accentset{(2)}{\tau}^{A}\wedge (\accentset{(0)}{\psi}^{+})^{t}C\gamma_{5}\gamma_{A}\wedge\accentset{(0)}{R}^{+}(Q)+\accentset{(0)}{\tau}^{A}\wedge (\accentset{(1)}{\psi}^{-})^{t}C\gamma_{5}\gamma_{A}\wedge\accentset{(1)}{R}^{-}(Q)+\nonumber\\
&+\accentset{(1)}{E}^{a}\wedge (\accentset{(0)}{\psi}^{+})^{t}C\gamma_{5}\gamma_{a}\wedge\accentset{(1)}{R}^{-}(Q)+\accentset{(1)}{E}^{a}\wedge (\accentset{(1)}{\psi}^{-})^{t}C\gamma_{5}\gamma_{a}\wedge\accentset{(0)}{R}^{+}(Q)\bigg]\ . \label{eq:actionp1}
\end{align}
\subsubsection{Membrane Case p=2}
In this case for the Rarita-Schwinger term we obtain
\begin{align}
\overline{\psi}\gamma_{5}\gamma_{\hat{A}}\wedge E^{\hat{A}}\wedge R(Q)=&(\psi^{+})^{t}C\gamma_{5}\gamma_{A}\wedge\tau^{A}\wedge R^{-}(Q)+(\psi^{-})^{t}C\gamma_{5}\gamma_{A}\wedge\tau^{A}\wedge R^{+}(Q)+\nonumber\\
&+(\psi^{+})^{t}C\gamma_{5}\gamma_{3}\wedge E\wedge R^{+}(Q)+(\psi^{-})^{t}C\gamma_{5}\gamma_{3}\wedge E\wedge R^{-}(Q)\ ,
\end{align}
where $E=E^{3}$. Then for the full Lagrangian we can write
\begin{align}
\mathcal{L}=\epsilon_{ABC}\bigg[&R^{AB}(J)\wedge \tau^{C} \wedge E+2R^{A}(G)\wedge \tau^{B} \wedge \tau^{C}\bigg]+\nonumber\\
+4\bigg[&(\psi^{+})^{t}C\gamma_{5}\gamma_{A}\wedge\tau^{A}\wedge R^{-}(Q)+(\psi^{-})^{t}C\gamma_{5}\gamma_{A}\wedge\tau^{A}\wedge R^{+}(Q)+\nonumber\\
&+(\psi^{+})^{t}C\gamma_{5}\gamma_{3}\wedge E\wedge R^{+}(Q)+(\psi^{-})^{t}C\gamma_{5}\gamma_{3}\wedge E\wedge R^{-}(Q)\bigg]\ .
\end{align}
Expanding at the lowest orders we have
\begin{subequations}
\begin{align}
\accentset{(1)}{\mathcal{L}}&=\epsilon_{ABC}\biggl[
\accentset{(0)}{R}^{AB}(J)\wedge \accentset{(0)}{\tau}^{A} \wedge \accentset{(1)}{E}+2\accentset{(1)}{R}^{A}(G)\wedge \accentset{(0)}{\tau}^{B} \wedge \accentset{(0)}{\tau}^{C}\biggr]+\nonumber\\
&-4\biggl[\accentset{(0)}{\tau}^{A}\wedge(\accentset{(0)}{\psi}^{+})^{t}C\gamma_{5}\gamma_{A}\accentset{(1)}{R}^{-}(Q)+\accentset{(0)}{\tau}^{A}\wedge(\accentset{(1)}{\psi}^{-})^{t}C\gamma_{5}\gamma_{A}\accentset{(0)}{R}^{+}(Q)+\accentset{(1)}{E}\wedge(\accentset{(0)}{\psi}^{+})^{t}C\gamma_{5}\gamma_{3}\accentset{(0)}{R}^{+}(Q)\biggr]\\
\accentset{(3)}{\mathcal{L}}&=\epsilon_{ABC}\biggl[\accentset{(0)}{R}^{AB}(J)\wedge \accentset{(0)}{\tau}^{A} \wedge \accentset{(3)}{E}+\accentset{(2)}{R}^{AB}(J)\wedge \accentset{(0)}{\tau}^{A} \wedge \accentset{(1)}{E}+\accentset{(0)}{R}^{AB}(J)\wedge \accentset{(2)}{\tau}^{A} \wedge \accentset{(1)}{E}+\nonumber\\
&+2\accentset{(3)}{R}^{A}(G)\wedge \accentset{(0)}{\tau}^{B} \wedge \accentset{(0)}{\tau}^{C}+4\accentset{(1)}{R}^{A}(G)\wedge \accentset{(2)}{\tau}^{B} \wedge \accentset{(0)}{\tau}^{C}\biggr]+\nonumber\\
&-4\biggl[\accentset{(0)}{\tau}^{A}\wedge(\accentset{(0)}{\psi}^{+})^{t}C\gamma_{5}\gamma_{A}\wedge\accentset{(3)}{R}^{-}(Q)+
\accentset{(2)}{\tau}^{A}\wedge(\accentset{(0)}{\psi}^{+})^{t}C\gamma_{5}\gamma_{A}\wedge\accentset{(1)}{R}^{-}(Q)+
\accentset{(0)}{\tau}^{A}\wedge(\accentset{(2)}{\psi}^{+})^{t}C\gamma_{5}\gamma_{A}\wedge\accentset{(1)}{R}^{-}(Q)+\nonumber\\
&\accentset{(0)}{\tau}^{A}\wedge(\accentset{(3)}{\psi}^{-})^{t}C\gamma_{5}\wedge\gamma_{A}\accentset{(0)}{R}^{+}(Q)+
+\accentset{(2)}{\tau}^{A}\wedge(\accentset{(1)}{\psi}^{-})^{t}C\gamma_{5}\gamma_{A}\wedge\accentset{(0)}{R}^{+}(Q)+
\accentset{(0)}{\tau}^{A}\wedge(\accentset{(1)}{\psi}^{-})^{t}C\gamma_{5}\gamma_{A}\wedge\accentset{(2)}{R}^{+}(Q)+\nonumber\\
&+\accentset{(3)}{E}\wedge(\accentset{(0)}{\psi}^{+})^{t}C\gamma_{5}\gamma_{3}\wedge\accentset{(0)}{R}^{+}(Q)+
\accentset{(1)}{E}\wedge(\accentset{(2)}{\psi}^{+})^{t}C\gamma_{5}\gamma_{3}\wedge\accentset{(0)}{R}^{+}(Q)+
\accentset{(1)}{E}\wedge(\accentset{(0)}{\psi}^{+})^{t}C\gamma_{5}\gamma_{3}\wedge\accentset{(2)}{R}^{+}(Q)+\nonumber\\
&+\accentset{(1)}{E}\wedge(\accentset{(1)}{\psi}^{-})^{t}C\gamma_{5}\gamma_{3}\wedge\accentset{(1)}{R}^{-}(Q)\biggr]\ .
\end{align}
\end{subequations}
For the same reasons considered in the previous subsection our action is given by the order three term,
\begin{align}
\mathcal{L}^{p=2}&=\epsilon_{ABC}\biggl[\accentset{(0)}{R}^{AB}(J)\wedge \accentset{(0)}{\tau}^{A} \wedge \accentset{(3)}{E}+\accentset{(2)}{R}^{AB}(J)\wedge \accentset{(0)}{\tau}^{A} \wedge \accentset{(1)}{E}+\accentset{(0)}{R}^{AB}(J)\wedge \accentset{(2)}{\tau}^{A} \wedge \accentset{(1)}{E}+\nonumber\\
&+2\accentset{(3)}{R}^{A}(G)\wedge \accentset{(0)}{\tau}^{B} \wedge \accentset{(0)}{\tau}^{C}+4\accentset{(1)}{R}^{A}(G)\wedge \accentset{(2)}{\tau}^{B} \wedge \accentset{(0)}{\tau}^{C}\biggr]+\nonumber\\
&-4\biggl[\accentset{(0)}{\tau}^{A}\wedge(\accentset{(0)}{\psi}^{+})^{t}C\gamma_{5}\gamma_{A}\wedge\accentset{(3)}{R}^{-}(Q)+
\accentset{(2)}{\tau}^{A}\wedge(\accentset{(0)}{\psi}^{+})^{t}C\gamma_{5}\gamma_{A}\wedge\accentset{(1)}{R}^{-}(Q)+
\accentset{(0)}{\tau}^{A}\wedge(\accentset{(2)}{\psi}^{+})^{t}C\gamma_{5}\gamma_{A}\wedge\accentset{(1)}{R}^{-}(Q)+\nonumber\\
&\accentset{(0)}{\tau}^{A}\wedge(\accentset{(3)}{\psi}^{-})^{t}C\gamma_{5}\wedge\gamma_{A}\accentset{(0)}{R}^{+}(Q)+
+\accentset{(2)}{\tau}^{A}\wedge(\accentset{(1)}{\psi}^{-})^{t}C\gamma_{5}\gamma_{A}\wedge\accentset{(0)}{R}^{+}(Q)+
\accentset{(0)}{\tau}^{A}\wedge(\accentset{(1)}{\psi}^{-})^{t}C\gamma_{5}\gamma_{A}\wedge\accentset{(2)}{R}^{+}(Q)+\nonumber\\
&+\accentset{(3)}{E}\wedge(\accentset{(0)}{\psi}^{+})^{t}C\gamma_{5}\gamma_{3}\wedge\accentset{(0)}{R}^{+}(Q)+
\accentset{(1)}{E}\wedge(\accentset{(2)}{\psi}^{+})^{t}C\gamma_{5}\gamma_{3}\wedge\accentset{(0)}{R}^{+}(Q)+
\accentset{(1)}{E}\wedge(\accentset{(0)}{\psi}^{+})^{t}C\gamma_{5}\gamma_{3}\wedge\accentset{(2)}{R}^{+}(Q)+\nonumber\\
&+\accentset{(1)}{E}\wedge(\accentset{(1)}{\psi}^{-})^{t}C\gamma_{5}\gamma_{3}\wedge\accentset{(1)}{R}^{-}(Q)\biggr]\ .\label{eq:actionp2}
\end{align}

\subsection{Invariance and Transformation Rules}
In this subsection we discuss the ivariance of the action under the algebra $\mathfrak{g}(2,3)$ that we have obtained. We focus on P-type symmetries, i.e. the transformations descending from space and time translations after the expansion and under supersymmetry since it is immediate to recognize that the transformations descending from Lorentz generators leave the action invariant (see also \cite{Bergshoeff:2019ctr} for further details).  

\subsubsection{P-type Symmetries}
The action is not invariant under P-type symmetries, i.e. the transformations generated by $P_{\hat{A}}$ (space and time translations), however we could apply an argument similar to that described in \cite{Bergshoeff:2019ctr} to see that this is not an independent symmetry and to derive the modified P-transformations leaving the action invariant up to a term proportional to the equation of motion of the spin connection. The action is invariant under Lorentz, general coordinate transformations and a trivial symmetry, or equation of motion symmetry; these last transforms the fields in terms proportional to the equation of motion. In particular if we write the variation as 
\begin{align}
\delta\mathcal{L}=A^{\mu}_{\hat{A}}\delta E^{\hat{A}}_{\mu}+B^{\mu}_{\hat{A}\hat{B}}\delta \Omega^{\hat{A}\hat{B}}_{\mu}+\overline{C}^{\mu}_{\alpha}\delta\psi^{\alpha}_{\mu}\ ,
\end{align}
where explicitly
\begin{subequations}
\begin{align}
A^{\rho}_{\hat{C}}&=E\left[R(J)E^{\rho}_{\hat{C}}-2R^{\rho}_{\hat{C}}(J)\right]+\epsilon^{\mu\nu\lambda\rho}\overline{\psi}_{\lambda}\gamma_{5}\gamma_{\hat{C}} R_{\mu\nu}(Q)\\
B^{\mu}_{\hat{A}\hat{B}}&=E\left[2E^{\mu}_{[\hat{A}}R^{\hat{C}}_{\hat{B}]\hat{C}}(P)+E^{\mu}_{\hat{C}}R_{\hat{A}\hat{B}}^{\hat{C}}(P)\right]\\
\overline{C}^{\lambda}_{\alpha}&=\epsilon^{\mu\nu\lambda\sigma}\left[2\gamma_{5}\gamma_{\sigma}R_{\mu\nu}(Q)+R_{\mu\sigma}^{\hat{A}}(P)\gamma_{5}\gamma_{\hat{A}}\psi_{\nu}\right]\ .
\end{align}
\end{subequations}
We recognize, among others,  the following trivial symmetry
\begin{subequations}
\begin{align}
\delta E_{\mu}^{\hat{A}}&=2ER^{\hat{A}}_{\mu\nu}(P)\sigma^{\nu}\\
\delta \Omega_{\mu}^{\hat{A}\hat{B}}&=-2A_{\mu}^{[\hat{A}}\sigma^{\hat{B}]}-E_{\mu}^{[\hat{B}}\sigma^{\hat{A}]}A^{\hat{C}}_{\hat{C}}+E_{\mu}^{[\hat{B}}A^{\hat{A}]}_{\hat{C}}\sigma^{\hat{C}}\\
\delta \psi_{\mu}^{\alpha}&=0\ .
\end{align}
\end{subequations}
In \autoref{sec:trivialsymmetry} we describe in details the trivial symmetries. The vielbein is the only field transforming under P-type symmetries.  The full transformations of the vielbein, apart from P-type transformation are given by the trivial symmetry. Lorentz, supersymmetry and general coordinate transformations 
\begin{align}
\delta E_{\mu}^{\hat{A}}&=2ER^{\hat{A}}_{\mu\nu}(P)\sigma^{\nu}-\lambda^{\hat{A}\hat{B}}E_{\mu\hat{B}}+\xi^{\nu}\partial_{\nu}E_{\mu}^{\hat{A}}+\partial_{\mu}\xi^{\nu}E_{\nu}^{\hat{A}}+i\overline{\epsilon}\gamma^{\hat{A}}\psi_{\mu}\ .
\end{align}
By choosing the parameters as 
\begin{multieq}[3]
\xi^{\nu}&=2E\sigma^{\nu}\\
\lambda^{\hat{A}\hat{B}}&=-\Omega^{\hat{A}\hat{B}}_{\nu}\xi^{\nu}\\
\epsilon&=\xi^{\nu}\psi_{\nu}\ ,
\end{multieq}
the transformation above reduces to
\begin{align}
\delta E_{\mu}^{\hat{A}}&=\partial_{\mu}\xi^{\hat{A}}+\Omega^{\hat{A}\hat{B}}\xi_{\hat{B}}\ ,
\end{align}
where $\xi^{\hat{A}}=E^{\hat{A}}_{\nu}\xi^{\nu}$. We recognize this as the P-type transformation with parameter $\eta^{\hat{A}}=\xi^{\hat{A}}$. Thus the P-type symmetries are not independent and could be written as a combination of supersymmetry, Lorentz , general coordinate transformations and the trivial symmetry. Rewriting the P-type transformation as combination of the other symmetries require the a modification of the P-type transformation of the gravitino and of the spin connection as follows
\begin{subequations}
\begin{align}
\delta E_{\mu}^{\hat{A}}&=\partial_{\mu}\xi^{\hat{A}}+\Omega^{\hat{A}\hat{B}}\xi_{\hat{B}}\\
\delta\psi_{\mu}&=\left(\partial_{\mu}+\frac{1}{4}\Omega_{\mu}^{\hat{A}\hat{B}}\gamma_{\hat{A}\hat{B}}\right)(\eta^{\hat{C}}\psi_{\hat{C}})+\frac{1}{4}\eta^{\nu}\Omega^{\hat{A}\hat{B}}_{\nu}\gamma_{\hat{A}\hat{B}}\psi_{\mu}\\
\delta \Omega_{\mu}^{\hat{A}\hat{B}}&=\frac{1}{2E}\left[-2A_{\mu}^{[\hat{A}}\eta^{\hat{B}]}-E_{\mu}^{[\hat{B}}\eta^{\hat{A}]}A^{\hat{C}}_{\hat{C}}+E_{\mu}^{[\hat{B}}A^{\hat{A}]}_{\hat{C}}\eta^{\hat{C}}\right]-\partial_{\mu}(\Omega_{\hat{C}}^{\hat{B}}\eta^{\hat{C}})-2\Omega_{\mu\phantom{A}\hat{C}}^{[\hat{A}}\Omega^{\hat{B}]\hat{C}}_{\hat{D}}\eta^{\hat{D}}\ .
\end{align}\label{eq:P-typemodified}
\end{subequations}
Since of the transformations we have used to write the P-type transformations only supersymmetry leave the action not invariant, this means that the invariance of the action is directly related to the behavior under supersymmetry that we are going to discuss in the next section. In particular putting the spin connection on-shell guarantees the invariance under supersymmetry that in turn, for the argument just exposed, implies the invariance under \autoref{eq:P-typemodified}.

\subsubsection{Supersymmetry}
The Lagrangian we have considered\cite{Deser:1976eh, CHAMSEDDINE197739}\ ,
\begin{align}
S=\epsilon_{\hat{A}\hat{B}\hat{C}\hat{D}}R^{\hat{A}\hat{B}}(J)\wedge E^{\hat{C}}\wedge E^{\hat{D}}+4\overline{\psi}\gamma_{5}\gamma_{\hat{A}}\wedge E^{\hat{A}}\wedge D\psi\ ,
\end{align}
is invariant under the supersymmetry transformations listed in \autoref{sec:generalcase} provided that the following constraint is imposed 
\begin{align}
R^{\hat{A}}(P)=0,
\end{align}
i.e. the spin connection is put on-shell. This means that the actions that we have found are invariant under the following supersymmetry transformations, for the $p=1$ and $p=2$ cases, when the corresponding constraints are satisfied, i.e. when the fields coming from the spin connection through the expansion, are put on-shell.
We can treat at once the two cases not specifying the values of the indices $A$ and $a$. It will be understood that certain terms will appear only if allowed by the specific case, for example $\Omega^{ab}$ exists for $p=1$ while it does not exist for $p=2$. The supersymmetry transformation rules are
\begin{subequations}
\begin{align}
\delta \accentset{(1)}{E}^{a}&=i(\accentset{(0)}{\epsilon}^{+})^{t}C\gamma^{a}\wedge\accentset{(1)}{\psi}^{-}+i(\accentset{(1)}{\epsilon}^{-})^{t}C\gamma^{a}\wedge\accentset{(0)}{\psi}^{+}\\
\delta \accentset{(3)}{E}^{a}&=i(\accentset{(0)}{\epsilon}^{+})^{t}C\gamma^{a}\wedge\accentset{(3)}{\psi}^{-}+i(\accentset{(3)}{\epsilon}^{-})^{t}C\gamma^{a}\wedge\accentset{(0)}{\psi}^{+}+i(\accentset{(2)}{\epsilon}^{+})^{t}C\gamma^{a}\wedge\accentset{(1)}{\psi}^{-}+i(\accentset{(1)}{\epsilon}^{-})^{t}C\gamma^{a}\wedge\accentset{(2)}{\psi}^{+}\\
\delta\accentset{(0)}{\tau}^{A}&=i(\accentset{(0)}{\epsilon}^{+})^{t}C\gamma^{A}\wedge\accentset{(0)}{\psi}^{+}\\
\delta\accentset{(2)}{\tau}^{A}&=i(\accentset{(2)}{\epsilon}^{+})^{t}C\gamma^{A}\wedge\accentset{(0)}{\psi}^{+}+i(\accentset{(0)}{\epsilon}^{+})^{t}C\gamma^{A}\wedge\accentset{(2)}{\psi}^{+}+i(\accentset{(1)}{\epsilon}^{-})^{t}C\gamma^{A}\wedge\accentset{(1)}{\psi}^{-}\\
\delta \accentset{(0)}{\psi}^{+}&=d\accentset{(0)}{\epsilon}^{+}-\frac{1}{4}\accentset{(0)}{\Omega}^{AB}\gamma_{AB}\accentset{(0)}{\epsilon}^{+}-\frac{1}{4}\accentset{(0)}{\Omega}^{ab}\gamma_{ab}\accentset{(0)}{\epsilon}^{+}\\
\delta \accentset{(2)}{\psi}^{+}&=d\accentset{(2)}{\epsilon}^{+}-\frac{1}{4}\accentset{(2)}{\Omega}^{AB}\gamma_{AB}\accentset{(0)}{\epsilon}^{+}-\frac{1}{4}\accentset{(0)}{\Omega}^{AB}\gamma_{AB}\accentset{(2)}{\epsilon}^{+}-\frac{1}{4}\accentset{(2)}{\Omega}^{ab}\gamma_{ab}\accentset{(0)}{\epsilon}^{+}-\frac{1}{4}\accentset{(0)}{\Omega}^{ab}\gamma_{ab}\accentset{(2)}{\epsilon}^{+}-\frac{1}{2}\accentset{(1)}{\Omega}^{Ab}\gamma_{Ab}\accentset{(1)}{\epsilon}^{-}\\
\delta\accentset{(1)}{\psi}^{-}&=d\accentset{(1)}{\epsilon}^{-}-\frac{1}{4}\accentset{(0)}{\Omega}^{AB}\gamma_{AB}\accentset{(1)}{\epsilon}^{-}-\frac{1}{4}\accentset{(0)}{\Omega}^{ab}\gamma_{ab}\accentset{(1)}{\epsilon}^{-}-\frac{1}{2}\accentset{(1)}{\Omega}^{Ab}\gamma_{Ab}\accentset{(0)}{\epsilon}^{+}\\
\delta\accentset{(3)}{\psi}^{-}&=d\accentset{(3)}{\epsilon}^{-}-\frac{1}{4}\accentset{(0)}{\Omega}^{AB}\gamma_{AB}\accentset{(3)}{\epsilon}^{-}\frac{1}{4}\accentset{(2)}{\Omega}^{AB}\gamma_{AB}\accentset{(1)}{\epsilon}^{-}-\frac{1}{4}\accentset{(0)}{\Omega}^{ab}\gamma_{ab}\accentset{(3)}{\epsilon}^{-}-\frac{1}{4}\accentset{(2)}{\Omega}^{ab}\gamma_{ab}\accentset{(1)}{\epsilon}^{-}-\frac{1}{2}\accentset{(3)}{\Omega}^{Ab}\gamma_{Ab}\accentset{(0)}{\epsilon}^{+}-\frac{1}{2}\accentset{(1)}{\Omega}^{Ab}\gamma_{Ab}\accentset{(2)}{\epsilon}^{+}\ 
\end{align}
\end{subequations}
while the curvature constraints read
\begin{subequations}
\begin{align}
\accentset{(1)}{R}^{a}(P)=&d\accentset{(1)}{E}^{a}+\accentset{(0)}{\Omega}^{ab}\wedge \accentset{(1)}{E}_{b}-\accentset{(1)}{\Omega}^{Aa}\wedge \accentset{(0)}{\tau}_{A}
-i(\accentset{(0)}{\psi}^{+})^{t}C\gamma^{a}\wedge\accentset{(1)}{\psi}^{-}=0\\
\accentset{(3)}{R}^{a}(P)=&d\accentset{(3)}{E}^{a}+\accentset{(2)}{\Omega}^{ab}\wedge \accentset{(1)}{E}_{b}
+\accentset{(0)}{\Omega}^{ab}\wedge \accentset{(3)}{E}_{b}-\accentset{(1)}{\Omega}^{Aa}\wedge \accentset{(2)}{\tau}_{A}
-\accentset{(3)}{\Omega}^{Aa}\wedge \accentset{(0)}{\tau}_{A}-i(\accentset{(0)}{\psi}^{+})^{t}C\gamma^{a}\wedge\accentset{(3)}{\psi}^{-}
-i(\accentset{(2)}{\psi}^{+})^{t}C\gamma^{a}\wedge\accentset{(1)}{\psi}^{-}
=0\\
\accentset{(0)}{R}^{A}(H)=&d\accentset{(0)}{\tau}^{A}+\accentset{(0)}{\Omega}^{AB}\wedge \accentset{(0)}{\tau}_{B}-\frac{i}{2}(\accentset{(0)}{\psi}^{+})^{t}C\gamma^{A}\wedge\accentset{(0)}{\psi}^{+}=0\\
\accentset{(2)}{R}^{A}(H)=&d\accentset{(2)}{\tau}^{A}+\accentset{(2)}{\Omega}^{AB}\wedge \accentset{(0)}{\tau}_{B}+\accentset{(0)}{\Omega}^{AB}\wedge \accentset{(2)}{\tau}_{B}+\accentset{(1)}{\Omega}^{Aa}\wedge \accentset{(1)}{E}_{a}-i(\accentset{(2)}{\psi}^{+})^{t}C\gamma^{A}\wedge\accentset{(0)}{\psi}^{+}-\frac{i}{2}(\accentset{(1)}{\psi}^{-})^{t}C\gamma^{A}\wedge\accentset{(1)}{\psi}^{-}=0\ .
\end{align}
\end{subequations}
What we have obtained is a set of two actions, one describing a non-relativistic supersymmetric string theory, the other non-relativistic supersymmetric theory of membranes. These action are invariant under the expanded supersymmetry transformations up to terms proportional to the expanded curvature $R(P)$. We do not investigate further the actions that we have obtained, postponing it to future works.

\section*{Discussion and Conclusions}\addcontentsline{toc}{section}{Discussion and Conclusions} 
In this work we have applied the procedure of Lie algebra expansion to the $\mathcal{N}=1$ super-Poincaré algebra in four dimensions, generalizing the results of \cite{deAzcarraga:2019mdn} for the $\mathcal{N}=2$ $p=0$ three dimensional super-Poincaré algebra. \\

In \autoref{sec:generalsetting} we have defined the general setting for the application of the Lie algebra expansion method. In particular we have considered the $\mathcal{N}=1$ $p$-brane super-Poincaré algebra in $D$-dimensions, $\mathfrak{g}$. We have then defined a decomposition in subspaces $\mathfrak{g}=V_{0}\oplus V_{1}$. In order to obtain a non-relativistic algebra after the expansion, and to be able to apply the Lie algebra expansion method, we have required this decomposition to satisfy two requirements, the first is that it induces a symmetric space structure, the second is that the Inonu-Wigner contraction of $\mathfrak{g}$ with respect to the subalgebra $V_{0}$ should be the $p$-brane super-Galilei algebra . The decomposition preparatory to the expansion method has required a splitting of the spinors using two sets of $p$-brane projectors. These projectors have been defined in \autoref{appendix:susyalgb}, where we describe in detail their application to the superlagebra and their properties. The projectors provide a completely general tool that could be immediately applied or straightforwardly generalized to different cases, including different signatures and $\mathcal{N}>1$. We have realized that, using this approach, it is possible to define a decomposition of the $\mathcal{N}=1$ super-Poincaré algebra such that we obtain two non-relativistic algebras for $p=1,2$ while for particle the algebra obtained is ultra-relativistic.\\
In \autoref{sec:N=1 D=4} we have applied the expansion procedure to the $\mathcal{N}=1$ four dimensional super-Poincaré algebra focusing on the action. Expanding the gauge fields associated to the expanded $p$-brane super-Poincaré algebra we have obtained two actions,  for strings and membranes, describing  two non-relativistic supersymmetric  theories. The actions that we find contain Fermion fields that have half the number of  degrees of freedom of four dimensional Majorana spinors.\\

The most natural extension of the present results is the application of the Lie algebra expansion method, trough the formalism we provide, to the ultra-relativistic case. It is also interesting to use our setting to study $\mathcal{N}>1$ supersymmetric theories. Furthermore it could be really interesting to inspect different superalgebra decomposition and to try to generalize the definition of the projectors.

\section*{Acknowledgments}
This work and LR have been supported in part by the
MINECO/FEDER, UE grant PGC2018-095205-B-I00 and by the Spanish Research Agency
(Agencia Estatal de Investigación) through the grant IFT Centro de Excelencia Severo Ochoa SEV-2016-0597. LR would like to thank José Manuel Izquierdo, Eric Bergshoeff and Tomás Ortín for precious discussions. 

\newpage
\appendix

\section{Supersymmetry Algebra and p-Brane Projectors}\label{appendix:susyalgb}
In this section we discuss how to act on the super-Poincaré algebra  $\mathfrak{g}$ in order to decompose it in way that is useful for the Lie algebra expansion. In particular what we want to obtain is a decomposition in two subspaces $\mathfrak{g}=V_{0}\oplus V_{1}$  such that they respect the following symmetric space structure 
\begin{multieq}[3]
[V_{0},V_{0}]&\subseteq V_{0}\\
[V_{0},V_{1}]&\subseteq V_{1}\\
[V_{1},V_{1}]&\subseteq V_{0}\ ,
\end{multieq}
where $[\cdot,\cdot]$ denotes the supercommutator. Furthermore since $\mathfrak{g}(0,1)$ will be the Inonu-Wigner contraction of $\mathfrak{g}$ with respect to $V_{0}$ and we are interested in obtaining, through the expansion, a non-relativistic algebra we would like to choose $V_{0}$ such that $\mathfrak{g}(0,1)$ is the superGaliliei algebra. By this way the higher order  expanded algebras $\mathfrak{g}(N_{0},N_{1})$ will be an extension of the $p$-brane super-Galilei algebras.\\
  
We consider a $D$ dimensional space with signature $(+,\ -,\ \cdots \ ,\ -)$ and a flat index decomposition $\hat{A}=\{A,a\}$, in general not related to the signature, with the index $A$ taking $p+1$ values, say $A=0,1,...,p$, and the index $a$ the remaining $D-p-1$ values. In this case we can always choose the gamma matrices such that $\gamma_{0}^{\dag}=\gamma_{0}$ and $\gamma_{i}^{\dag}=-\gamma_{i}$ for $i=1, ..., D-1$. We define the following matrices
\begin{align}
U&=\alpha \gamma_{0}\gamma_{1}...\gamma_{p}\\
V&=\beta \gamma_{p+1}...\gamma_{D-1}\ , 
\end{align} 
where $\alpha$ and $\beta$ are two phase factors, and see how they act on the gamma matrices. We note that 
\begin{align}
V&=\pm\gamma_{*}U\ .
\end{align}
where $\gamma_{*}=\eta\gamma_{0}...\gamma_{D-1}$ and $\eta$ is a phase factor. These matrices satisfy
\begin{multieq}[2]
U^{\dag}&=(\alpha^{*})^2(-1)^{\frac{p(p+3)}{2}}U\\
U^{\dag}U&=UU^{\dag}=1\\
V^{\dag}&=(\beta^{*})^2(-1)^{\frac{(D-p-1)(D-p)}{2}}V\\
V^{\dag}V&=VV^{\dag}=1
\end{multieq}
and
\begin{multieq}[2]
U\gamma_{A}U^{\dag}&=(-1)^{p\phantom{+1}}\ \gamma_{A}\\
U\gamma_{a}U^{\dag}&=(-1)^{p+1}\ \gamma_{a}\\
V\gamma_{A}V^{\dag}&=(-1)^{D-p-1}\ \gamma_{A}\\
V\gamma_{a}V^{\dag}&=(-1)^{D-p\phantom{-1}}\ \gamma_{a}\ .
\end{multieq}

We have thus built two matrices that act discriminating the two subsets of our index splitting $\hat{A}=\{A,a\}$. We can then define the following projectors
\begin{align}
\Pi^{U}_{\pm}&=\frac{1}{2}\left(1\pm U\right)\\
\Pi^{V}_{\pm}&=\frac{1}{2}\left(1\pm V\right)\ ,\label{eq:projectorsdefinition}
\end{align}
but in order for this to be projectors we should require $UU=1$ and $VV=1$ that will fix the phase factors to 
\begin{multieq}[2]
\alpha&=(-1)^{-\frac{p(p+3)}{4}}\\
\beta&=(-1)^{-\frac{(D-p)(D-p-1)}{4}}\ .
\end{multieq}
With these choices we immediately recognize that
\begin{multieq}[2]
U^{\dag}&=U\\
V^{\dag}&=V.
\end{multieq}
The definition of the projectors induces the following definitions
\begin{subequations}
\begin{align}
Q^{\pm}=\Pi^{U}_{\pm}Q
\end{align}
or
\begin{align}
Q^{\pm}=\Pi^{V}_{\pm}Q\ . 
\end{align}
\end{subequations}
The Dirac adjoint is defined by
\begin{align}
\overline{\psi}=\psi^{\dag}\mathcal{D}\ ,
\end{align} 
where we consider $\mathcal{D}$ as in \cite{ortin_2004}
\begin{align}
\mathcal{D}=\delta \gamma_{0}
\end{align}
and $\delta$ is a phase factor. This matrix satisfies
\begin{subequations}
\begin{align}
U\mathcal{D}&=(-1)^{p}\mathcal{D}U\\
V\mathcal{D}&=(-1)^{D-p-1}\mathcal{D}V\ ,
\end{align}
\end{subequations}
implying , in the two cases
\begin{subequations}
\begin{align}
\overline{Q}^{\pm}&=\overline{(\Pi_{\pm}^{U}Q)}=\overline{Q}\Pi_{\pm(-1)^{p}}^{U}\\
\overline{Q}^{\pm}&=\overline{(\Pi_{\pm}^{V}Q)}=\overline{Q}\Pi_{\pm(-1)^{D-p-1}}^{V}\ .
\end{align}
\end{subequations}
Then the commutation relations of the super-Poincaré algebra involving the Fermionic generator
\begin{subequations}
\begin{align}
\{Q^{\alpha},\overline{Q}_{\beta}\}&=k(\gamma^{\hat{A}})^{\alpha}{}_{\beta}P_{\hat{A}}\\
[J_{\hat{A}\hat{B}},Q^{\alpha}]&=-\frac{1}{2}\left(\gamma_{\hat{A}\hat{B}}\right)^{\alpha}{}_{\beta}Q^{\beta}\\
[J_{\hat{A}\hat{B}},\overline{Q}_{\alpha}]&=+\frac{1}{2}\left(\overline{Q}\gamma_{\hat{A}\hat{B}}\right)_{\alpha}
\end{align}
\end{subequations}
applying the projector $\Pi^{U}_{\pm}$ and $\Pi^{V}_{\pm}$, become 
\begin{multicols}{2}
\setlength{\columnseprule}{0.4pt}
\centering {$\mathbf{\Pi_{\pm}^{U}}$}
\begin{subequations}
\begin{align}
\{Q^{\alpha\pm},\overline{Q}^{\pm}_{\beta}\}&=k\left[\Pi^{U}_{\pm}\gamma_{A}\right]^{\alpha}{}_{\beta}H^{A}\\
\{Q^{\alpha\pm},\overline{Q}^{\mp}_{\beta}\}&=k\left[\Pi^{U}_{\pm}\gamma_{a}\right]^{\alpha}{}_{\beta}P^{a}\\
[J_{AB},Q^{\alpha\pm}]&=-\frac{1}{2}\left(\gamma_{AB}\right)^{\alpha}{}_{\beta}Q^{\beta\pm}\\
[J_{ab},Q^{\alpha\pm}]&=-\frac{1}{2}\left(\gamma_{ab}\right)^{\alpha}{}_{\beta}Q^{\beta\pm}\\
[G_{Ab},Q^{\alpha\pm}]&=-\frac{1}{2}\left(\gamma_{Ab}\right)^{\alpha}{}_{\beta}Q^{\beta\mp}
\end{align}\label{eq:algebraN=2}
\end{subequations}
\centering {$\mathbf{\Pi_{\pm}^{V}}$}
\begin{subequations}
\begin{align}
\{Q^{\alpha\pm},\overline{Q}^{\pm}_{\beta}\}&=k\left[\Pi^{V}_{\pm}\gamma_{A}\right]^{\alpha}{}_{\beta}H^{A}\\
\{Q^{\alpha\pm},\overline{Q}^{\mp}_{\beta}\}&=k\left[\Pi^{V}_{\pm}\gamma_{a}\right]^{\alpha}{}_{\beta}P^{a}\\
[J_{AB},Q^{\alpha\pm}]&=-\frac{1}{2}\left(\gamma_{AB}\right)^{\alpha}{}_{\beta}Q^{\beta\pm}\\
[J_{ab},Q^{\alpha\pm}]&=-\frac{1}{2}\left(\gamma_{ab}\right)^{\alpha}{}_{\beta}Q^{\beta\pm}\\
[G_{Ab},Q^{\alpha\pm}]&=-\frac{1}{2}\left(\gamma_{Ab}\right)^{\alpha}{}_{\beta}Q^{\beta\mp}\ .
\end{align}
\end{subequations}
\end{multicols}

Now we collect some useful relations before we discuss the case of Majorana spinors,
\begin{multieq}[2]
\Pi_{\pm}^{U}\gamma_{*}&=\gamma_{*}\Pi_{\pm(-1)^{(p+1)(D+1)}}^{U}\\
\Pi_{\pm}^{U}\gamma_{A}&=\gamma_{A}\Pi_{\pm(-1)^{p}}^{U}\\
\Pi_{\pm}^{U}\gamma_{a}&=\gamma_{a}\Pi_{\pm(-1)^{p+1}}^{U}\\
(\Pi_{\pm}^{U})^{\dag}&=\Pi_{\pm}^{U}\\
\Pi_{\pm}^{V}\gamma_{*}&=\gamma_{*}\Pi_{\pm(-1)^{(D+1)(D-p-1)}}^{V}\\
\Pi_{\pm}^{V}\gamma_{A}&=\gamma_{A}\Pi_{\pm(-1)^{D-p-1}}^{V}\\
\Pi_{\pm}^{V}\gamma_{a}&=\gamma_{a}\Pi_{\pm(-1)^{D-p}}^{V}\\
(\Pi_{\pm}^{V})^{\dag}&=\Pi_{\pm}^{V}\ .
\end{multieq}
For a complex spinor
\begin{multieq}[2]
\psi^{\pm}&=\Pi_{\pm}^{U}\psi\\
(\psi^{\pm})^{\dag}&=\psi^{\dag}\Pi_{\pm}^{U}\\
\overline{\psi}^{\pm}&=\overline{\psi}\Pi_{\pm(-1)^{p}}^{U}\\
\psi^{\pm}&=\Pi_{\pm}^{V}\psi\\
(\psi^{\pm})^{\dag}&=\psi^{\dag}\Pi_{\pm}^{V}\\
\overline{\psi}^{\pm}&=\overline{\psi}\Pi_{\pm(-1)^{D-p-1}}^{V}\ .
\end{multieq}
This implies, considering the following bilinears, with the two different projectors, that we can write
\begin{description}
\item[$\mathbf{\Pi_{\pm}^{U}}$]
\begin{subequations}
\begin{align}
\overline{\psi}\gamma_{A_{1}}...\gamma_{A_{i}}\gamma_{a_{1}}...\gamma_{a_{j}}\chi&=\overline{\psi}^{+}\gamma_{A_{1}}...\gamma_{A_{i}}\gamma_{a_{1}}...\gamma_{a_{j}}\chi^{+(-1)^{p(i+j+1)+j}}+\nonumber\\
&+\overline{\psi}^{-}\gamma_{A_{1}}...\gamma_{A_{i}}\gamma_{a_{1}}...\gamma_{a_{j}}\chi^{-(-1)^{p(i+j+1)+j}}\\
\overline{\psi}\gamma_{*}\gamma_{A_{1}}...\gamma_{A_{i}}\gamma_{a_{1}}...\gamma_{a_{j}}\chi&=\overline{\psi}^{+}\gamma_{*}\gamma_{A_{1}}...\gamma_{A_{i}}\gamma_{a_{1}}...\gamma_{a_{j}}\chi^{+(-1)^{p(i+j+1)+j+(p+1)(D-1)}}+\nonumber\\
&+\overline{\psi}^{-}\gamma_{*}\gamma_{A_{1}}...\gamma_{A_{i}}\gamma_{a_{1}}...\gamma_{a_{j}}\chi^{-(-1)^{p(i+j+1)+j+(p+1)(D-1)}}
\end{align}
\end{subequations}

\item[$\mathbf{\Pi_{\pm}^{V}}$]
\begin{subequations}
\begin{align}
\overline{\psi}\gamma_{A_{1}}...\gamma_{A_{i}}\gamma_{a_{1}}...\gamma_{a_{j}}\chi&=\overline{\psi}^{+}\gamma_{A_{1}}...\gamma_{A_{i}}\gamma_{a_{1}}...\gamma_{a_{j}}\chi^{+(-1)^{(D-p-1)(i+j+1)+j}}+\nonumber\\
&+\overline{\psi}^{-}\gamma_{A_{1}}...\gamma_{A_{i}}\gamma_{a_{1}}...\gamma_{a_{j}}\chi^{-(-1)^{(D-p-1)(i+j+1)+j}}\\
\overline{\psi}\gamma_{*}\gamma_{A_{1}}...\gamma_{A_{i}}\gamma_{a_{1}}...\gamma_{a_{j}}\chi&=\overline{\psi}^{+}\gamma_{*}\gamma_{A_{1}}...\gamma_{A_{i}}\gamma_{a_{1}}...\gamma_{a_{j}}\chi^{+(-1)^{(D-p-1)(D+i+j)+j}}+\nonumber\\
&+\overline{\psi}^{-}\gamma_{*}\gamma_{A_{1}}...\gamma_{A_{i}}\gamma_{a_{1}}...\gamma_{a_{j}}\chi^{-(-1)^{(D-p-1)(D+i+j)+j}}\ .
\end{align}
\end{subequations}
\end{description}

\subsection{Majorana Spinor }\label{sec:Majorana}

Now we consider the case in which $Q$ is a Majorana spinor. This means
\begin{align}
\overline{Q}=Q^{C}=Q^{t}C_{(x)}\ ,
\end{align}
where $C_{(x)}$ is the similarity matrix (charge conjugation matrix) defined by
\begin{align}
C_{(x)}\gamma_{\hat{A}}C^{-1}_{(x)}=x\gamma^{t}_{\hat{A}}
\end{align}
and $x=\pm$. The Majorana conjugate spinor $Q^{C}$ is in components
\begin{align}
Q^{C}_{\alpha}=Q^{\beta}(C_{(x)})_{\beta\alpha}\ .
\end{align}
In this case we can find a basis such that all the gamma matrices are purely imaginary and from now on we adopt this as our basis. We stress that not both $C_{(\pm)}$ exist for any dimension and furthermore that , when they both exist the consistency of the Majorana condition, and thus the supersymmetry algebra, require a precise choice. The argument that we present in this section is completely general but , due to the remarks just done should be carefully applied.\\  
We can write
\begin{subequations}
\begin{align}
C_{(x)}\gamma_{0}C^{-1}_{(x)}&=-x\gamma_{0}\\
C_{(x)}\gamma_{i}C^{-1}_{(x)}&=+x\gamma_{i}\ ,
\end{align}
\end{subequations}
for $i=1,...,D-1$. Thus we have
\begin{multieq}[2]
C_{(x)}UC^{-1}_{(x)}&=x^{p}U\\
C_{(x)}VC^{-1}_{(x)}&=x^{D-p-1}V\ ,
\end{multieq}

that is
\begin{multieq}[2]
C_{(x)}U&=x^{p}UC_{(x)}\\
C_{(x)}V&=x^{D-p-1}VC_{(x)}\ .
\end{multieq}
The projectors thus satisfy
\begin{multieq}[2]
C_{(x)}\Pi_{\pm}^{U}&=\Pi_{\pm x^{p}}^{U}C_{(x)}\\
C_{(x)}\Pi_{\pm}^{V}&=\Pi_{\pm x^{D-p-1}}^{U}C_{(x)}\ ,
\end{multieq}
then in the two cases we have
\begin{subequations}
\begin{alignat}{6}
&\overline{Q}^{\pm}&&=\overline{Q}\Pi_{\pm(-1)^{p}}^{U}&&=Q^{t}C_{(x)}\Pi_{\pm (-1)^{p}}^{U}&&=Q^{t}\Pi_{\pm(-1)^{p}x^{p}}^{U}C_{(x)}\\
&\overline{Q}^{\pm}&&=\overline{Q}\Pi_{\pm(-1)^{D-p-1}}^{V}&&=Q^{t}C_{(x)}\Pi_{\pm(-1)^{D-p-1}}^{U}&&=Q^{t}\Pi_{\pm(-1)^{D-p-1}x^{D-p-1}}^{V}C_{(x)}\ .
\end{alignat}
\end{subequations}
Since
\begin{subequations}
\begin{align}
U^{t}&=+(-1)^{\frac{p(p+1)}{2}+1}U\\
V^{t}&=+(-1)^{\frac{(D-p-2)(D-p-1)}{2}}V\ ,
\end{align}
\end{subequations}
we get
\begin{multicols}{2}
\setlength{\columnseprule}{0.4pt}
\centering {$\mathbf{\Pi_{\pm}^{U}}$}
\begin{subequations}
\begin{align}
\overline{Q}^{\pm}&=\left(Q^{\pm (-1)^{\frac{(p+1)(p+2)}{2}} x^{p}}\right)^{t}C_{(x)}
\end{align}
\end{subequations}
\centering {$\mathbf{\Pi_{\pm}^{V}}$}
\begin{subequations}
\begin{align}
\overline{Q}^{\pm}&=\left(Q^{\pm (-1)^{\frac{(D-p)(D-p-1)}{2}} x^{D-p-1}}\right)^{t}C_{(x)}\ .
\end{align}
\end{subequations}
\end{multicols}
In particular, using $C=C_{(-)}$, 
\begin{multicols}{2}
\setlength{\columnseprule}{0.4pt}
\centering {$\mathbf{\Pi_{\pm}^{U}}$}
\begin{subequations}
\begin{align}
\overline{Q}^{\pm}&=\left(Q^{\pm (-1)^{\frac{p(p+1)}{2}+1}}\right)^{t}C
\end{align}
\end{subequations}
\centering {$\mathbf{\Pi_{\pm}^{V}}$}
\begin{subequations}
\begin{align}
\overline{Q}^{\pm}&=\left(Q^{\pm (-1)^{\frac{(D-p-1)(D-p+2)}{2}}}\right)^{t}C\ .
\end{align}
\end{subequations}
\end{multicols}
We can thus define
\begin{subequations}
\begin{align}
\xi^{U}&=(-1)^{\frac{p(p+1)}{2}+1}\\
\xi^{V}&=(-1)^{\frac{(D-p-1)(D-p+2)}{2}}\ .
\end{align}
\end{subequations}
Finally  the anticommutators of the algebra, using the two different projectors, could be put in the form
\begin{multicols}{2}
\setlength{\columnseprule}{0.4pt}
\centering {$\mathbf{\Pi_{\pm}^{U}}$}
\begin{subequations}
\begin{align}
\{Q^{\alpha\pm},Q^{\beta\pm\xi^{U} }\}&=k\left[\Pi^{U}_{\pm}\gamma_{A}C^{-1}\right]^{\alpha\beta}H^{A}\\
\{Q^{\alpha\pm}, Q^{\beta\mp \xi^{U}}\}&=k\left[\Pi^{U}_{\pm}\gamma_{a}C^{-1}\right]^{\alpha\beta}P^{a}
\end{align}
\end{subequations}
\centering {$\mathbf{\Pi_{\pm}^{V}}$}
\begin{subequations}
\begin{align}
\{Q^{\alpha\pm},Q^{\beta\pm \xi^{V}}\}&=k\left[\Pi^{V}_{\pm}\gamma_{A}C^{-1}\right]^{\alpha\beta}H^{A}\\*
\{Q^{\alpha\pm},Q^{\beta\mp \xi^{V}}\}&=k\left[\Pi^{V}_{\pm}\gamma_{a}C^{-1}\right]^{\alpha\beta}P^{a}\ .
\end{align}
\end{subequations}
\end{multicols}
By this way we can recognize that the interesting case for our purpose is that with $\xi^{U}=+1$, i.e. $p=1,2$. In these cases the  decomposition of the superalgebra satisfying our requests, with the use of the projectors $\Pi_{\pm}^{U}$ is $V_{0}=\{J_{AB},\ J_{ab}, H^{A},\ Q^{+} \}$. We note that the case $\xi^{U}=-1$ corresponds to the ultra-relativistic case; we will not treat this case in the present work.\\
We specialize now the spinor bilinear discussed in the previous section to the case in which they are Majorana. We have using $\Pi_{\pm}^{U}$ or $\Pi_{\pm}^{V}$
\begin{description}
\item[$\mathbf{\Pi_{\pm}^{U}}$]
\begin{subequations}
\begin{align}
\overline{\psi}\gamma_{A_{1}}...\gamma_{A_{i}}\gamma_{a_{1}}...\gamma_{a_{j}}\chi&=\left(\psi^{+(-1)^{\frac{p(p+1)}{2}+1}}\right)^{t}C\gamma_{A_{1}}...\gamma_{A_{i}}\gamma_{a_{1}}...\gamma_{a_{j}}\chi^{+(-1)^{p(i+j+1)+j}}+\nonumber\\
&+\left(\psi^{-(-1)^{\frac{p(p+1)}{2}+1}}\right)^{t}C\gamma_{A_{1}}...\gamma_{A_{i}}\gamma_{a_{1}}...\gamma_{a_{j}}\chi^{-(-1)^{p(i+j+1)+j}}\\
\overline{\psi}\gamma_{*}\gamma_{A_{1}}...\gamma_{A_{i}}\gamma_{a_{1}}...\gamma_{a_{j}}\chi&=\left(\psi^{+(-1)^{\frac{p(p+1)}{2}+1}}\right)^{t}C\gamma_{*}\gamma_{A_{1}}...\gamma_{A_{i}}\gamma_{a_{1}}...\gamma_{a_{j}}\chi^{+(-1)^{p(i+j+1)+j+(p+1)(D-1)}}+\nonumber\\
&+\left(\psi^{-(-1)^{\frac{p(p+1)}{2}+1}}\right)^{t}C\gamma_{*}\gamma_{A_{1}}...\gamma_{A_{i}}\gamma_{a_{1}}...\gamma_{a_{j}}\chi^{-(-1)^{p(i+j+1)+j+(p+1)(D-1)}}
\end{align}
\end{subequations}

\item[$\mathbf{\Pi_{\pm}^{V}}$]
\begin{subequations}
\begin{align}
\overline{\psi}\gamma_{A_{1}}...\gamma_{A_{i}}\gamma_{a_{1}}...\gamma_{a_{j}}\chi&=\left(\psi^{+(-1)^{\frac{(D-p-1)(D-p+2)}{2}}}\right)^{t}C\gamma_{A_{1}}...\gamma_{A_{i}}\gamma_{a_{1}}...\gamma_{a_{j}}\chi^{+(-1)^{(D-p-1)(i+j+1)+j}}+\nonumber\\
&+\left(\psi^{-(-1)^{\frac{(D-p-1)(D-p+2)}{2}}}\right)^{t}C\gamma_{A_{1}}...\gamma_{A_{i}}\gamma_{a_{1}}...\gamma_{a_{j}}\chi^{-(-1)^{(D-p-1)(i+j+1)+j}}\\
\psi\gamma_{*}\gamma_{A_{1}}...\gamma_{A_{i}}\gamma_{a_{1}}...\gamma_{a_{j}}\chi&=\left(\psi^{+(-1)^{\frac{(D-p-1)(D-p+2)}{2}}}\right)^{t}C\gamma_{*}\gamma_{A_{1}}...\gamma_{A_{i}}\gamma_{a_{1}}...\gamma_{a_{j}}\chi^{+(-1)^{(D-p-1)(D+i+j)+j}}+\nonumber\\
&+\left(\psi^{-(-1)^{\frac{(D-p-1)(D-p+2)}{2}}}\right)^{t}C\gamma_{*}\gamma_{A_{1}}...\gamma_{A_{i}}\gamma_{a_{1}}...\gamma_{a_{j}}\chi^{-(-1)^{(D-p-1)(D+i+j)+j}}\ .
\end{align}
\end{subequations}
\end{description}

In particular in this last case we specify to $D=4$, $i=1,j=0$ and $i=0,j=1$ we get respectively for the two projectors
\begin{description}
\item[$\mathbf{\Pi_{\pm}^{U}}$]
\begin{subequations}
\begin{align}
\overline{\psi}\gamma_{*}\gamma_{A}\chi&=\left(\psi^{+(-1)^{\frac{p(p+1)}{2}+1}}\right)^{t}C\gamma_{*}\gamma_{A}\chi^{+(-)^{p+1}}+\left(\psi^{-(-1)^{\frac{p(p+1)}{2}+1}}\right)^{t}C\gamma_{*}\gamma_{A}\chi^{-(-)^{p+1}}\\
\overline{\psi}\gamma_{*}\gamma_{a}\chi&=\left(\psi^{+(-1)^{\frac{p(p+1)}{2}+1}}\right)^{t}C\gamma_{*}\gamma_{a}\chi^{+(-)^{p}}+\left(\psi^{-(-1)^{\frac{p(p+1)}{2}+1}}\right)^{t}C\gamma_{*}\gamma_{a}\chi^{-(-)^{p}}
\end{align}
\end{subequations}

\item[$\mathbf{\Pi_{\pm}^{V}}$]
\begin{subequations}
\begin{align}
\overline{\psi}\gamma_{*}\gamma_{A}\chi&=\left(\psi^{+(-1)^{\frac{(3-p)(6-p)}{2}}}\right)^{t}C\gamma_{*}\gamma_{A}\chi^{+(-)^{p+1}}+\left(\psi^{-(-1)^{\frac{(3-p)(6-p)}{2}}}\right)^{t}C\gamma_{*}\gamma_{A}\chi^{-(-)^{p+1}}\\
\overline{\psi}\gamma_{*}\gamma_{a}\chi&=\left(\psi^{+(-1)^{\frac{(3-p)(6-p)}{2}}}\right)^{t}C\gamma_{*}\gamma_{a}\chi^{+(-)^{p}}+\left(\psi^{-(-1)^{\frac{(3-p)(6-p)}{2}}}\right)^{t}C\gamma_{*}\gamma_{a}\chi^{-(-)^{p}}\ .
\end{align}
\end{subequations}
\end{description}

\subsection{Consistency Condition for Majorana}
We consider  the consistency condition in four dimensions ($C$ is real for our choice of basis) between Majorana and Dirac adjoint spinors 
\begin{align}
\overline{\psi}=\psi^{c}\ ,
\end{align}
that is
\begin{align}
\psi^{\dag}\mathcal{D}=\psi^{t}C_{(x)}.
\end{align}
This gives
\begin{align}
\mathcal{D}^{t}\psi^{*}=C_{(x)}^{t}\psi\ .
\end{align}
In particular in our case, since $\mathcal{D}^{t}=-\mathcal{D}$ and $C_{(x)}^{t}=-C$  and $C_{(x)}^{2}=-1$ in four dimensions, we obtain
\begin{align}
\psi= C_{(x)}\mathcal{D}\psi^{*}\label{eq:Majoranaconsistency1}\ .
\end{align}
Taking the complex conjugate we have (note $\mathcal{D}\mathcal{D}^{*}=-1$)
\begin{align}
\psi=\mathcal{D}C_{(x)}\psi^{*}
\end{align}
from which we get 
\begin{align}
\psi=-xC_{(x)}\mathcal{D}\psi^{*}\label{eq:Majoranaconsistency2}\ . 
\end{align}
Comparing \autoref{eq:Majoranaconsistency1} with \autoref{eq:Majoranaconsistency2} we immediately recognize in four dimensions the consistent choice is $C_{(-)}$.

\section{Useful relations}
In this section we list some relations useful in studying the invariance of the action and the derivation of the equation of motion.
\begin{subequations}
\begin{align}
\gamma_{\hat{A}}\gamma_{\hat{B}}\gamma_{\hat{C}}&=\eta_{\hat{A}\hat{B}}\gamma_{\hat{C}}+\eta_{\hat{B}\hat{C}}\gamma_{\hat{A}}-\eta_{\hat{A}\hat{C}}\gamma_{\hat{B}}-i\epsilon_{\hat{A}\hat{B}\hat{C}\hat{D}}\gamma^{\hat{D}}\gamma_{5}\\
\gamma_{\hat{C}\hat{D}}\gamma_{\hat{A}\hat{B}}&=2\eta_{\hat{A}\hat{D}}\gamma_{\hat{C}\hat{B}}-2\eta_{\hat{A}\hat{C}}\gamma_{\hat{D}\hat{B}}+2\eta_{\hat{D}\hat{B}}\gamma_{\hat{A}\hat{C}}-2\eta_{\hat{C}\hat{B}}\gamma_{\hat{A}\hat{D}}+\gamma_{\hat{A}\hat{B}}\gamma_{\hat{C}\hat{D}}\\
\gamma_{\hat{A}\hat{B}}\gamma_{\hat{C}}&=2\eta_{\hat{B}\hat{C}}\gamma_{\hat{A}}-2\eta_{\hat{A}\hat{C}}\gamma_{\hat{B}}+\gamma_{\hat{C}}\gamma_{\hat{A}\hat{B}}\\
\gamma_{\sigma} \gamma_{\hat{A}\hat{B}} &= \gamma_{\sigma\hat{A}\hat{B}} + 2E_{\sigma[\hat{A}}\gamma_{\hat{B}]} =iE^{\hat{D}}_{\sigma}\epsilon_{\hat{A}\hat{B}\hat{C}\hat{D}}\gamma_{*}\gamma^{\hat{C}}+2E_{\sigma[\hat{A}}\gamma_{\hat{B}]}\\
\overline{\psi}\gamma_{\hat{A}}\chi&=-\overline{\chi}\gamma_{\hat{A}}\psi\\
\overline{\psi}\gamma_{5}\gamma_{\hat{A}}\chi&=\overline{\chi}\gamma_{5}\gamma_{\hat{A}}\psi\\
\overline{\psi}\gamma_{5}\gamma_{\hat{A}}\gamma_{\hat{B}\hat{C}}\chi&=-\overline{\chi}\gamma_{5}\gamma_{\hat{B}\hat{C}}\gamma_{\hat{A}}\psi\ .
\end{align}
\end{subequations}
The variation uder susy in differential form language
\begin{align*}
\delta S=2R^{\hat{A}}(P)\wedge\overline{\epsilon}\gamma_{5}\gamma_{\hat{A}}\wedge R(Q)
\end{align*}

\section{Trivial Symmetry}\label{sec:trivialsymmetry}
In this section we study the trivial symmetries of the $\mathcal{N}=1$ supergravity action in four dimensions. The trivial symmetries or equation of motion symmetries are define by the transformations of the fields to be proportional to the equations of motion. As a starting point we take the action in the following form
\begin{align}
\mathcal{L}=EE^{\mu}_{\hat{A}}E^{\nu}_{\hat{B}}R^{\hat{A}\hat{B}}_{\mu\nu}(J)+2\epsilon^{\mu\nu\lambda\sigma}\overline{\psi}_{\lambda}\gamma_{5}\gamma_{\sigma} D_{\mu}\psi_{\nu}
\end{align}
and we compute the variations with respect to the three fields, obtaining 
\begin{subequations}
\begin{align}
\delta_{E}\mathcal{L}&=E\left[R(J)E^{\rho}_{\hat{C}}-2R^{\rho}_{\hat{C}}(J)\right]\delta E_{\rho}^{\hat{C}}+\epsilon^{\mu\nu\lambda\rho}\overline{\psi}_{\lambda}\gamma_{5}\gamma_{\hat{C}} R_{\mu\nu}(Q)
\delta E_{\rho}^{\hat{C}}\\
\delta_{\Omega}\mathcal{L}&=2\epsilon_{\hat{A}\hat{B}\hat{C}\hat{D}}R^{\hat{A}}(P)\wedge E^{\hat{B}}\wedge \delta\Omega^{\hat{C}\hat{D}}\nonumber\\
&=E\left[E^{\mu}_{\hat{A}}E^{\nu}_{\hat{B}}E^{\rho}_{\hat{C}}
+E^{\mu}_{\hat{B}}E^{\nu}_{\hat{C}}E^{\rho}_{\hat{A}}+E^{\mu}_{\hat{C}}E^{\nu}_{\hat{A}}E^{\rho}_{\hat{B}}\right]R_{\nu\rho}^{\hat{C}}(P)\delta \Omega_{\mu}^{\hat{A}\hat{B}}\nonumber\\
&=E\left[2E^{\mu}_{[\hat{A}}R^{\hat{C}}_{\hat{B}]\hat{C}}(P)+E^{\mu}_{\hat{C}}R_{\hat{A}\hat{B}}^{\hat{C}}(P)\right]\delta\Omega_{\mu}^{\hat{A}\hat{B}}\\
\delta_{\psi}\mathcal{L}
&=\delta\overline{\psi}_{\lambda}\epsilon^{\mu\nu\lambda\sigma}\left[2\gamma_{5}\gamma_{\sigma}R_{\mu\nu}(Q)+R_{\mu\sigma}^{\hat{A}}(P)\gamma_{5}\gamma_{\hat{A}}\psi_{\nu}\right]\ .
\end{align}\label{eq:fieldvariantions1}
\end{subequations}

We denote the full variation of the Lagrangian as
\begin{align}
\delta\mathcal{L}=A^{\mu}_{\hat{A}}\delta E^{\hat{A}}_{\mu}+B^{\mu}_{\hat{A}\hat{B}}\delta \Omega^{\hat{A}\hat{B}}_{\mu}+\overline{C}^{\mu}_{\alpha}\delta\psi^{\alpha}_{\mu}\ ,
\end{align}
where $=A^{\mu}_{\hat{A}}$, $B^{\mu}_{\hat{A}\hat{B}}$ and $\overline{C}^{\mu}_{\alpha}$ could be immediately read from \autoref{eq:fieldvariantions1}. We will define three trivial symmetries to which we will refer as type $AB$, $BC$ and $AC$ from the equations of motion involved in the transformations.  

\subsection{Type AB}
We consider the following transformation
\begin{subequations}
\begin{align}
\delta E_{\mu}^{\hat{A}}&=a B_{\mu\hat{B}}^{\hat{A}}\sigma^{\hat{B}}+
b E_{\mu}^{\hat{A}}B_{\hat{B}\hat{C}}^{\hat{C}}\sigma^{\hat{B}}+cB_{\mu\hat{B}}^{\hat{B}}\sigma^{\hat{A}}\\
\delta \Omega_{\mu}^{\hat{A}\hat{B}}&=-aA_{\mu}^{[\hat{A}}\sigma^{\hat{B}]}-bE_{\mu}^{[\hat{B}}\sigma^{\hat{A}]}A^{\hat{C}}_{\hat{C}}-cE_{\mu}^{[\hat{B}}A^{\hat{A}]}_{\hat{C}}\sigma^{\hat{C}}\\
\delta \psi_{\mu}^{\alpha}&=0\ ,
\end{align}
\end{subequations}
where $a,b,c$ are arbitrary constants. 
We note that taking $a=2, b=1, c=-1$ we have 
\begin{align}
\delta E_{\mu}^{\hat{A}}&=2 B_{\mu\hat{B}}^{\hat{A}}\sigma^{\hat{B}}+
E_{\mu}^{\hat{A}}B_{\hat{B}\hat{C}}^{\hat{C}}\sigma^{\hat{B}}-B_{\mu\hat{B}}^{\hat{B}}\sigma^{\hat{A}}=2ER^{\hat{A}}_{\mu\nu}(P)\sigma^{\nu}\ .
\end{align}

\subsection{Type BC}
For the type BC trivial symmetry we have the follwing transformations
\begin{subequations}
\begin{align}
\delta E_{\mu}^{\hat{A}}&=0\\
\delta\Omega^{\hat{A}\hat{B}}_{\mu}&=-aE_{\mu}^{[\hat{B}}\chi_{\nu}^{\hat{A}]}C^{\nu}-bC^{[\hat{A}}E_{\mu}^{\hat{B}]}\chi^{\hat{C}}_{\hat{C}}-cC^{[\hat{A}}\chi_{\mu}^{\hat{B}]}\\
\delta\psi_{\mu}^{\alpha}&=a\chi_{\mu}^{\alpha\hat{A}}B_{\hat{A}\hat{B}}^{\hat{B}}+b\chi^{\hat{A}\alpha}_{\hat{A}}B_{\mu\hat{B}}^{\hat{B}}+c\chi^{\hat{A}\alpha}_{\hat{B}}B_{\mu\hat{A}}^{\hat{B}}\ .
\end{align}
\end{subequations}
With $c=2,a=1,b=-1$ we have
\begin{align}
\delta \psi_{\mu}^{\alpha}=2ER_{\mu\nu}^{\hat{B}}(P)\chi^{\nu\alpha}_{\hat{B}}\ .
\end{align}

\subsection{Type AC}
Finally for the type AC we have
\begin{subequations}
\begin{align}
\delta E_{\mu}^{\hat{A}}&=a\overline{C}^{A}\zeta_{\mu}+bE_{\mu}^{\hat{A}}\overline{C}^{\hat{B}}\zeta_{\hat{B}}\\
\delta\Omega^{\hat{A}\hat{B}}_{\mu}&=0\\
\delta\psi_{\mu}^{\alpha}&=-a\zeta^{\alpha}_{\hat{A}}A^{\hat{A}}_{\mu}-bA^{\hat{A}}_{\hat{A}}\zeta_{\mu}^{\alpha}\ .
\end{align}
\end{subequations}
If we set $a=2,b=-1$ we get
\begin{align}
\delta\psi_{\mu}^{\alpha}=4ER_{\mu}^{\hat{A}}(J)\zeta_{A}^{\alpha}\ .
\end{align}

\subsection{Lorentz + General Coordiante Transformations }

Under Lorentz and general coordinate transformations we have
\begin{subequations}
\begin{align}
\delta E^{\hat{A}}_{\mu}&=-\lambda^{\hat{A}\hat{B}}E_{\mu\hat{B}}+\xi^{\nu}\partial_{\nu}E_{\mu}^{\hat{A}}+\partial_{\mu}\xi^{\nu}E_{\nu}^{\hat{A}}\\
\delta \Omega_{\mu}^{\hat{A}\hat{B}}&=\partial_{\mu}\lambda^{\hat{A}\hat{B}}
-2\Omega^{[\hat{A}}_{\mu\phantom{A}\hat{C}}\lambda^{\hat{B}]\hat{C}}
+\xi^{\nu}\partial_{\nu}\Omega_{\mu}^{\hat{A}\hat{B}}+\partial_{\mu}\xi^{\nu}\Omega_{\nu}^{\hat{A}\hat{B}}\\
\delta \psi^{\alpha}_{\mu}&=-\frac{1}{4}\lambda^{\hat{A}\hat{B}}\gamma_{\hat{A}\hat{B}}\psi_{\mu}+\xi^{\nu}\partial_{\nu}\psi_{\mu}+\partial_{\mu}\xi^{\nu}\psi_{\nu}\ .
\end{align}
\end{subequations}

\section{Equations of Motion of the Spin Connection and Expansion}
In this section we list the equations of motion of the fields coming from the spin connection after the expansion for the actions \autoref{eq:actionp1} and \autoref{eq:actionp2}. The equation of motion for the spin connection before the expansion and splitting of the indices is given by
\begin{align}
\delta_{\Omega}\mathcal{L}&=2\epsilon_{\hat{A}\hat{B}\hat{C}\hat{D}}R^{\hat{A}}(P)\wedge E^{\hat{B}}\wedge \delta\Omega^{\hat{C}\hat{D}}
\end{align}


The equations of motion for the fields coming from the spin connection in the two cases, $p=1,2$,   are
\begin{description}
\item[p=1]
\begin{subequations}
\begin{align}
\epsilon_{ABab}\ \delta\accentset{(0)}{\Omega}^{AB}\wedge \accentset{(1)}{R}^{a}(P)\wedge \accentset{(1)}{E}^{b}&=0\\
\epsilon_{ABab}\ \delta\accentset{(2)}{\Omega}^{ab}\wedge \accentset{(0)}{R}^{A}(H)\wedge \accentset{(0)}{\tau}^{B}&=0\\
\epsilon_{ABab}\ \delta\accentset{(0)}{\Omega}^{ab}\wedge\left[ \accentset{(2)}{R}^{A}(H)\wedge \accentset{(0)}{\tau}^{B}+\accentset{(0)}{R}^{A}(H)\wedge \accentset{(2)}{\tau}^{B}\right]&=0\\
\epsilon_{ABab}\ \delta\accentset{(1)}{\Omega}^{Aa}\wedge\left[ \accentset{(0)}{R}^{A}(H)\wedge \accentset{(1)}{E}^{b}-\accentset{(1)}{R}^{b}(P)\wedge \accentset{(0)}{\tau}^{B}\right]&=0\ .
\end{align}
\end{subequations}
Note that in the $p=1$ case the fields $\accentset{(3)}{\Omega}^{Ab}$ and $\accentset{(2)}{\Omega}^{AB}$ do not appear in the Lagrangian. 

\item[p=2]
\begin{subequations}
\begin{align}
\epsilon_{ABC}\ \delta\accentset{(2)}{\Omega}^{AB}\wedge\left[ \accentset{(0)}{R}^{C}(H)\wedge \accentset{(1)}{E}-\accentset{(1)}{R}(P)\wedge \accentset{(0)}{\tau}^{C}\right]&=0\\
\epsilon_{ABC}\ \delta\accentset{(0)}{\Omega}^{AB}\wedge\left[ \accentset{(2)}{R}^{C}(H)\wedge \accentset{(1)}{E}+\accentset{(0)}{R}^{C}(H)\wedge \accentset{(3)}{E}-\accentset{(3)}{R}(P)\wedge \accentset{(0)}{\tau}^{C}-\accentset{(1)}{R}(P)\wedge \accentset{(2)}{\tau}^{C}\right]&=0\\
\epsilon_{ABC}\ \delta\accentset{(1)}{\Omega}^{A}\wedge\left[ \accentset{(2)}{R}^{B}(H)\wedge \accentset{(0)}{\tau}^{C}+\accentset{(0)}{R}^{B}(H)\wedge \accentset{(2)}{\tau}^{C}\right]&=0\\
\epsilon_{ABC}\ \delta\accentset{(3)}{\Omega}^{A}\wedge \accentset{(0)}{R}^{B}(H)\wedge \accentset{(0)}{\tau}^{C}&=0\ .
\end{align}
\end{subequations}
\end{description}

\FloatBarrier

\end{document}